\documentclass[preprintnumbers,amsmath,amssymb,floatfix,10pt,prd,onecolumn,
superscriptaddress,nofootinbib]{revtex4-2}

\usepackage[parfill]{parskip}    
\usepackage{graphicx}
\usepackage{amssymb}
\usepackage{epstopdf}
\usepackage{color}
\usepackage{float}
\usepackage{amsmath}

\usepackage{graphicx}
\usepackage[colorlinks=true,citecolor=blue,urlcolor=magenta,breaklinks]{hyperref}

\begin{document}

\title{$f(R)$ wormholes embedded in a pseudo--Euclidean space $E^{5}$ }

\author{A. S. Agrawal}
\email{agrawalamar61@gmail.com}
\affiliation{Department of Mathematics, Birla Institute of Technology and Science-Pilani, Hyderabad Campus, Hyderabad 500078, India.}

\author{B. Mishra}
\email{bivu@hyderabad.bits-pilani.ac.in}
\affiliation{Department of Mathematics, Birla Institute of Technology and Science-Pilani, Hyderabad Campus, Hyderabad 500078, India.}

\author{Francisco Tello-Ortiz}
\email{francisco.tello@ua.cl}
\affiliation{Departamento de F\'isica, Facultad de Ciencias Básicas, Universidad de Antofagasta, Casilla 170, Antofagasta, Chile.}

\author{A. Alvarez}
\email{aalvarezu88@gmail.com}
\affiliation{Unidad de Equipamiento Cient\'ifico (Maini), Universidad Cat\'olica del Norte, 
Av. Angamos 0610, Antofagasta, Chile.}

\begin{abstract}
We introduce the\ldots
\end{abstract}
\begin{abstract}
This work is devoted to the study of analytic wormhole solutions within the framework of $f(R)$ gravity theory. To check the possibility of having wormhole structures satisfying energy conditions, by means of the class I approach the pair $\{\Phi(r), b(r)\}$ describing the wormhole geometry has been obtained. Then, in conjunction with a remarkably $f(R)$ gravity model, the satisfaction of the null and weak energy conditions at the wormhole throat and its neighborhood is investigated. To do so, 
some constant parameters have been bounded restricting the space parameter. In this concern, the $f(R)$ gravity model and its derivatives are playing a major role, specially in considering the violation of the non--existence theorem. Furthermore, the shape function should be bounded from above by the Gronwall--Bellman shape function, where the red--shift function plays a relevant role. By analyzing the main properties at the spatial stations and tidal accelerations at the wormhole throat, possibilities and conditions for human travel are explored.

\end{abstract}
\maketitle

\section{Introduction}\label{sec1}

Since the pioneering article by Morris and Thorne \cite{Morris88}, the study of wormhole space--times has compromised a lot of effort in understanding this kind of structures. In few words, a wormhole is a region formed by a curved surface of minimal size determined by $r_{0}$, the size of the throat, connecting two infinite or asymptotically flat regions \cite{Morris88,visser}. Interestingly, these types of objects are solutions of the general relativity (GR from now on) field equations. However, in the context of GR these structures exhibit a peculiar characteristic, which seems to be a general rule that must be satisfied for them to exist \cite{r13}. Specifically, such a characteristic concerns that the matter distribution supporting the wormhole corresponds to a particular type of matter, the so--called exotic matter \cite{Morris88,visser}. This type of matter is characterized by not respecting the energy conditions. Concretely the null energy condition (NEC) and weak energy condition (WEC) \cite{visser,curiel}. These conditions are the weakest of the energy conditions, when an observer moves at the speed of light it has well defined limit called the NEC $(\rho +p_i\geq 0)$ while the WEC $(\rho \geq0$ and $\rho +p_i\geq 0)$ says that the energy density of any system at any point of the space--time of any time--like observer is positive \cite{curiel}. Moreover, the NEC is being even weaker than the WEC \cite{curiel}.

Along these last three decades, dozens of articles devoted to the study of these kind of space--times have been published. For instance, in order to connect the wormhole cosmology and horizon problem,  the wormhole cosmology with Friedman--Robertson--Walker(FRW) space--time allowing two--way transmission of signals between spatially separated regions of space--time has been constructed \cite{Hochberg93}. Also, it has been shown that the WEC is satisfied when the Einstein field equations impose a contracting wormhole solution \cite{Arellano06}. The wormholes cannot be formed within the scenario of scalar–-tensor of dark energy admitting a phantom like behavior. \cite{Bronnikov07,Bro}. Besides, in $2+1$--dimensional gravity, it has been shown expanding wormholes for positive cosmological constant and expands to maximum and re--collapse for a negative cosmological constant \cite{Cataldo11}. The stability of the thin--shell wormholes geometry in dilaton gravity with generalized Chaplygin gas has been studied in \cite{Bejarano11}. In the context of third order Lovelock gravity, the higher--dimensional thin--shell wormholes has been studied in \cite{Mehdizadeh15} and it was shown that in the positive third order and negative second order Lovelock coefficient, wormhole solutions are satisfying the WEC. In the same line, in the context of scale--free $R^{2}$ gravity,
wormhole solutions are supported by the vacuum of the theory,
then there is not presence of exotic matter \cite{Duplessis15}. A Lorentzian wormhole space--time was obtained where the matter satisfies the WEC in scalar--tensor gravity \cite{Shaikh16}. Also, wormhole solutions have been studied with power law dependent and constant red--shift function \cite{Mehdizadeh17}. Employing the minimal coupling of anisotropic
matter field and vector field, a wormhole solution has been obtained
in \cite{Maldacena17}. The theoretical predictions for static wormhole solutions have been investigated and it has been suggested that the throat radius of the wormholes can be a constant \cite{Moraes17} and  the violation of NEC may attribute to the existence of the extra dimension \cite{Kuhfittig18}.   

Beyond the above, a plausible scenario for studying wormholes is the modified gravity theory $f(R)$ \cite{Buch70,Sta80,su2}. In this regard and in connection with the scalar--tensor theories this type of solutions are ruled out if they satisfy at the same time the tranversability and energy conditions. The only way of having a traversable wormhole satisfying the energy conditions on the background of the $f(R)$ gravity theory is by means of the propagation of states with negative kinetic energy i.e., the so--called ghost fields \cite{Bronnikov07,Bro}. In this respect, some authors have investigated and obtained interesting result in the research field. For example, conditions are imposed such that the matter threading the wormhole satisfies the energy conditions and obtained the wormhole geometry solution \cite{Lobo09}. The exact solution of static wormholes with the matter that contains the Lorentzian density distribution has been derived in \cite{Rahaman14} and the existence of stable static configuration with a suitable set of the model parameters has been shown in \cite{Eiroa18}. The cosmological wormhole solutions have been considered in \cite{Bahamonde16}. On the other hand, through the approach of Noether symmetry, the geometry of static traversable wormhole was studied in \cite{Sharif18} and the satisfaction of NEC near the wormhole throat of a flat and hyperbolic wormhole has been shown \cite{Golchin19}. A new class of f(R)--gravity models with wormhole solutions and cosmological properties have been obtained in \cite{Ortiz20}. Some more research on wormhole solutions in $f(R)$ gravity are presented in \cite{Pavlovic15,Mazharimousavi16,Godani20,Yousaf20,Tangphati20,Azad20,Shamir20}, what is more in \cite{Dai20} a simple description of the formation of wormholes by placing two massive objects in two parallel universes was given.  

As it is well--known, find out analytic solutions to the $f(R)$ gravity field equations is not an easy task \cite{su2}. This is so because the system of equations contains high order derivative terms in the metric variables, what is more, in the simplest case in dealing with a perfect or imperfect fluid distribution the system has four and five unknowns respectively, and only three equations. Then it is unavoidable to prescribe some additional information in to order to close the problem at least from the mathematical point of view. As we are interested in finding analytic wormhole solutions, we have implemented the so--called class I condition to derive the full geometry of the space--time \cite{eisland,eisenhart,karmarkar,sharma}, being described by the pair of metric functions $\{\Phi(r),b(r)\}$, the red--shift and shape function. The motivation behind this methodology is that the class I in the spherically symmetric case, relates both metric potentials. Therefore, given one of them the second one is immediately determined. Thus, this technology helps to reduce the number of unknowns in order to solve the system of equations. It is remarkable to mention that this techniques has been mainly used in the construction of stellar interiors \cite{r35,r36,r37,r39,r41,r43,Ramos21} (and references therein) and also used to obtain wormhole solutions in the GR arena \cite{K1,tello,K2,K3}. Additionally, we have assumed the $f(R)$ gravity model given in \cite{Tripathy16} to obtain the energy--momentum tensor and its components. To check the feasibility of the resulting models an exhaustive geometric analysis has been performed, also the satisfaction of the NEC and WEC at the wormhole throat and its neighborhood is assured by constraining some constant parameters. Respect to this, we have ensured the violation of the non--existence theorem by searching regions where $F(R)<0$ \cite{Bronnikov07,Bro}. Besides, a full analysis using the procedure given in \cite{Morris88,visser} is employed in determining under what conditions the present  model is satisfying the traversability criterion for human travel. 

The paper is organized as follows: Section \ref{sec2}, presents the main requirements that any wormhole space--time should satisfy. The Section \ref{sec3} presents the basic formalism and the field equations of the $f(R)$ gravity theory, adapted to the wormhole geometry. Section \ref{sec4} provides a short revisiting on the class I condition and also it is presented the model, its main features and energy conditions constraints are discussed. In section \ref{sec5.1}, traversability conditions for human trips are explored. Finally, Section \ref{sec6} concludes the present investigation. Throughout the article the mostly positive signature $-,+,+,+$ has been employed.

\section{Wormhole Taxonomy}\label{sec2}

In this section we present the general requirements that all geometrical structure should satisfy in order to represent a wormhole space--time. To do this, we shall follow the pioneering article by Morris and Thorne \cite{Morris88} and also the reference \cite{visser}. So, in canonical coordinates $x^{\mu}=(t,r,\theta,\phi)$, the most general line element describing a spherical and static wormhole is given by
\begin{equation}\label{eq1}
ds^2=-e^{\Phi}dt^2+\frac{1}{1-\frac{b}{r}}dr^2+r^2(d\theta^2+sin^2\theta d\phi^2),
\end{equation}
where the objects $\Phi$ and $b$ are purely radial functions i.e, $\Phi=\Phi(r)$ and $b=b(r)$. The former is the so--called red--shift function and the second one the form or shape function. In addition with the general line element (\ref{eq1}), to depict a wormhole space--time i.e., a throat or tunnel connecting two infinite or asymptotically flat space--times, both $\Phi$ and $b$ should meet the following constraints 
\begin{enumerate}
    \item The throat size connecting the regions is defined by a global minimum radius $r_{0}=\text{min\{r(l)\}}$, being $l$ the proper distant, covering the entire range $(-\infty,+\infty)$. Therefore, the radial coordinate $r$ covers the range $[r_{0},+\infty)$. The relationship between $l$ and $r$ is given by
    \begin{equation}
        l(r)=\pm \int^{r}_{r_{0}}\frac{dr}{\sqrt{1-\frac{b(r)}{r}}}.
    \end{equation}
     \item The red--shift function $\Phi$ must be finite everywhere, for $r\geq r_{0}$, in order to avoid an event horizon, so $e^{\Phi(r)}>0$ for all $r>r_{0}$. Moreover, 
    \begin{equation} \nonumber
        \lim_{r\rightarrow +\infty} \Phi(r)=\Phi_{0},
    \end{equation}
    where $\Phi_{0}$ is finite and real.
    \item The flare--out condition must hold
    \begin{equation}\nonumber
    \frac{b(r)-b^{\prime}(r)r}{2b^{2}(r)}>0,   
    \end{equation}
    at or near the throat $r=r_{0}$.
    \item The above condition implies that for all $r\geq r_{0}$: $b(r_{0})=r_{0}$ and $b^{\prime}(r_{0})< 1$. Furthermore, for all $r>r_{0}$ $\Rightarrow$ $b(r)<r$.
    \item To ensure the asymptotic behavior, the shape function $b(r)$ must fulfill
    \begin{equation} \nonumber
        \lim_{r\rightarrow +\infty} b(r)=\text{finite}.
    \end{equation}
    \item Taking into account the last point, the asymptotically flat behavior at enough large distances needs, 
    \begin{equation}\nonumber
        \lim_{r\rightarrow+\infty}\frac{b(r)}{r}\rightarrow 0.
    \end{equation}
\end{enumerate}
From now on primes will denote differentiation with respect to the radial coordinate $r$, that is, $\frac{d}{dr}\equiv \ '$.

\section{The $f(R)$ Field Equations}\label{sec3}

As mentioned in the introduction part, another type of modified theory of gravity that generalizes Einstein's General Relativity is the $f(R)$ gravity, being $R$ the Ricci scalar. The $f(R)$ gravity is a family of gravity theories where each one is defined by different function of the Ricci scalar. The most simplified function is when the function is equal to the scalar $R$, recovering GR. As we are intending to study wormhole geometries in the background of $f(R)$ theory of gravity, the starting point is the action
\begin{equation} \label{eq2}
S=\int\sqrt{-g}\left[\frac{f(R)}{2\kappa}+\mathcal{L}_m\right]d^4x,
\end{equation}
where $\kappa=8\pi G/c^{4}$, $\mathcal{L}_m$ being the matter Lagrangian density and $g$ the determinant of the metric $g_{\mu \nu}$. Hereinafter units where the coupling constant $\kappa$ is equal to one will be employed. Now, with the variation of the action with respect to $g^{\mu\nu}$ and incorporating the metric approach, one can obtain the general field equations of $f(R)$ gravity as,
\begin{equation} \label{eq3}
[R_{\mu \nu}-\nabla_{\mu}\nabla_{\nu}+g_{\mu \nu}\Box]F(R)-\frac{1}{2}fg_{\mu \nu}= T_{\mu \nu}
\end{equation}
where $F(R)$ is defined as follows 
\begin{equation}\label{eq4}
F(R)\equiv \frac{df(R)}{dR}.    
\end{equation}
Next, taking the trace of the Equation (\ref{eq3}) one gets
\begin{equation} \label{eq5}
RF(R)+3\square F(R)-2f(R)= T
\end{equation} 
where $T=g^{\mu \nu}T_{\mu \nu}$ is the trace of the energy--momentum tensor. Incorporating the above contraction and after some algebra \cite{Lobo09}, the $f(R)$ gravity field equations can be recast as, 

\begin{equation} \label{eq6}
G_{\mu \nu}=R_{\mu \nu}-\frac{1}{2}Rg_{\mu\nu}= T_{\mu\nu}^{EF},
\end{equation}
where $G_{\mu\nu}$ is the Einstein tensor.
The right member of Equation (\ref{eq6}) has been designed as the effective energy--momentum tensor $T_{\mu\nu}^{EF}=T^{(c)}_{\mu\nu}+\hat{T}^{(m)}_{\mu\nu}$, which is the combination of curvature energy--momentum tensor,
\begin{equation}\label{eq7}
T_{\mu \nu}^{(c)}\equiv\frac{1}{F} \left[\nabla_{\mu}\nabla_{\nu}-\frac{1}{4}(RF+\square F+T)g_{\mu \nu}\right]    
\end{equation}
and $ \widehat{T}_{\mu \nu}^{(m)}\equiv{{T}_{\mu \nu}^{(m)}}/{F(R)}$, where ${T}_{\mu \nu}^{(m)}$ is representing the energy--momentum tensor of the matter content. In this opportunity ${T}_{\mu \nu}^{(m)}$ is described by an imperfect matter distribution i.e., unequal stresses $p_{r}\neq p_{\perp}$. Specifically, the form of the energy--momentum tensor for this kind of matter content is given by,
\begin{equation}\label{eq8}
T^{(m)}_{\mu \nu}=(\rho+ p_{\perp})u_{\mu} u_{\nu} +(p_r-p_{\perp})\chi_{\mu}\chi_{\nu}+p_{\perp}g_{\mu \nu},
\end{equation} 
where $\chi^{\mu}$ is a space--like unit vector satisfying $\chi^{\mu}\chi_{\mu}=1$, $u^{\mu}$ a time--like normalized vector, representing the  four--velocity of the fluid satisfying $u^{\mu} u_{\mu}=-1$ and orthogonal to $\chi^{\mu}$, $\rho$ is denoting the energy density, $p_{r}$ and $p_\perp$ are the radial and transverse pressure, respectively. Putting together  Eqs. (\ref{eq6})--(\ref{eq8}) one arrives to the following expressions  \cite{Bahamonde16,Pavlovic15}
\begin{equation}\label{eq9}
\begin{aligned}
-\rho(r)=&-\frac{1}{2} f(r)+\left(1-\frac{b(r)}{r}\right) R^{\prime 2}(r) f_{R R R}(r)+\frac{F(r)}{2 r^{2} }\left[\left(r\left(b^{\prime}(r)-4\right)+3 b(r)\right) \Phi^{\prime}(r)-2 r(r-b(r)) \Phi^{\prime 2}(r)\right.\\
&\left.+2 r(b(r)-r) \Phi^{\prime \prime}(r)\right]+\frac{f_{R R}(r)}{2 r^{2} }\left[2 r(r-b(r)) R^{\prime \prime}(r)-\left(r\left(b^{\prime}(r)-4\right)+3 b(r)\right) R^{\prime}(r)\right] \\ 
 p_{r}(r)=&-\frac{1}{2} f(r)+\frac{b(r) F(r)}{2 r^{3} }\left[2 r^{2} \Phi^{\prime 2}(r)+2 r^{2} \Phi^{\prime \prime}(r)-r \Phi^{\prime}(r)-2\right]+\frac{F(r)}{2 r^{2}}\left[b^{\prime}(r)\left(r \Phi^{\prime}(r)+2\right)-2 r^{2}\left(\Phi^{\prime 2}(r)+\Phi^{\prime \prime}(r)\right)\right]\\
&+{f_{R R}(r) R^{\prime}}(r)\left(1-\frac{b(r)}{r}\right)\left(\Phi^{\prime}(r)+\frac{2}{r}\right) \\ 
 p_{\perp}(r)=&-\frac{1}{2} f(r)+\frac{F(r)}{2 r^{3} }\left[b(r)\left(2 r \Phi^{\prime}(r)+1\right)+\left(b^{\prime}(r)-2 r \Phi^{\prime}(r)\right) r\right]+f_{R R R}(r)R^{\prime 2}(r)\left(1-\frac{b(r)}{r}\right)\\
&+\frac{f_{R R}(r)}{2 r^{2}}\left[R^{\prime}(r)\left(r\left(2 r \Phi^{\prime}(r)-b^{\prime}(r)+2\right)-b(r)\left(2 r \Phi^{\prime}(r)+1\right)\right)+2 r(r-b(r)) R^{\prime \prime}(r)\right],
\end{aligned}
\end{equation}
where
\begin{equation}\label{eq12}
f_{RR}(r)\equiv f_{RR}(R(r))\equiv \frac{d^{2}f(R(r))}{dR^{2}(r)}, \quad   f_{RRR}(r)\equiv f_{RRR}(R(r))\equiv\frac{d^{3}f(R(r))}{dR^{3}(r)}.  
\end{equation}
To clarify the dependency of the variables and to avoid any confusion, it should be noted that the $f(R)$ gravity functional and all its derivatives, depend on the radial coordinate through the scalar curvature $R=R(r)$. Then, instead of write $F(R(r))$ in the expressions (\ref{eq9}), we have written just $F(r)$ and so on. Nevertheless, the derivative is with respect to $R$ and after plug the corresponding expression for $R$ in terms of the radial coordinate $r$, one is getting $f(r)$, $F(r)$, etc. This nomenclature shall be used along the manuscript.

Next, taking into account expressions (\ref{eq9}) one can get the so--called energy conditions, concretely the NEC and WEC, which their general form is given by \cite{visser}
\begin{align}\label{eq13}
\text{Null energy condition (NEC)}&: \rho+p_{r}\geq 0 \quad \mbox{and} \quad \rho+p_{\perp}\geq0, \\ \label{eq14} 
\text{Weak energy condition (WEC)} &: \rho\geq0,\quad \rho+p_{r}\geq 0 \quad \mbox{and} \quad \rho+p_{\perp}\geq0.
\end{align}
The main point in solving this tricky system of equations (\ref{eq9}) in dealing with wormhole--like structures, is the satisfaction at the same time of the geometric requirements listed in Sec. \ref{sec2} (specially the flare--out condition) and the so--called energy conditions, at least the NEC and WEC at the wormhole throat. As it is well--known, by construction, this situation is not possible within the arena of GR. Of course, as was pointed out by Hochberg and Visser \cite{r13}, Einstein field equations subject to the space--time (\ref{eq1}) do not allow to obtain positive defined thermodynamic quantities i.e., density $\rho$, radial $p_{r}$ and tangential $p_{\perp}$ pressure, violating in this way the energy conditions \cite{visser}, in this case the NEC and WEC. Although in general it is possible to get $\rho>0$ everywhere, but $p_{r}<0$ where $|p_{r}|>|\rho|$ then (\ref{eq13})--(\ref{eq14}) are violated in the radial direction. The above situation not only occurs in the context of GR, in this respect, scalar--tensor theories are facing some inconvenient to meet at the same footing the wormhole structure supported by normal matter (matter respecting the energy conditions) too. In this concern, Bronnikov and Starobinsky have established certain requirements under these conditions are simultaneously satisfied \cite{Bronnikov07,Bro}. In general a scalar--tensor theory is being described in the Jordan frame $\mathbb{M}_{J}$ by the following Lagrangian
\begin{equation}\label{eq15}
L=\frac{1}{2}\left[f(\Psi) R+h(\Psi) g^{\mu \nu} \partial_{\mu}\Psi \partial_{\nu}\Psi-2 U(\Psi)\right]+L_{m},
\end{equation}
where $f$, $h$ and $U$ are arbitrary functions and $L_{m}$ is the  Lagrangian of the matter sector. For this theory a static wormhole solution satisfying the energy conditions cannot exist if 
\begin{equation}\label{eq16}
f(\Psi)>0 \quad \mbox{and} \quad f(\Psi)h(\Psi)+\frac{3}{2}\left(\frac{df(\Psi)}{d\Psi}\right)^{2}>0.    
\end{equation}
The first statement of (\ref{eq16}) is related with the quantum stability of the theory. In this case this condition means that the graviton is not a ghost field whilst the second condition of (\ref{eq16}) implies that the scalar field $\Psi$ is not a ghost. Recalling that ghost fields correspond to states with negative kinetic energy leading to exponential instabilities. As scalar--tensor theories are related with $f(R)$ gravity under the following identification $h(\Psi)=0$ and $f(\Psi)=F(R)$ \cite{Bronnikov07}. The  analogous statements (\ref{eq16}) in this scenario are \cite{Bronnikov07,Bro}  
\begin{equation}\label{eq17}
F(R)=\frac{df(R)}{dR}>0 \quad \mbox{and} \quad \frac{d^{2}f(R)}{dR^{2}}\neq 0.    
\end{equation}
The conditions (\ref{eq16}) for the scalar--tensor theory or equivalently (\ref{eq17}) for the present case, $f(R)$ gravity theory, are known as the non--existence theorem of static wormhole solutions satisfying the energy conditions. Therefore, the only possibility to achieve the desired wormhole structure satisfying at least (\ref{eq13})--(\ref{eq14}) is by means of the violation of the theorem (\ref{eq17}). Thus, we shall look for regions where $F(R)<0$, then the theory propagates ghost fields, or $d^{2}f(R)/dR^{2}=0$. It is worth mentioning that, the violation of the first statement in (\ref{eq17}) provides a wide spectrum in the seeking of wormhole solutions since it is violated in a region at space, in contrast with the second condition which is violated only locally, that is, in one point. 

\section{The Class I Model}\label{sec4}

In this section, we revisited in short under what conditions a 4--dimensional space--time is classified as a class I space--time. Once this point is clear, we relate the class I approach with the general wormhole line element (\ref{eq1}) in order to establish the geometry of the wormhole manifold. In this concern, one should follow the guidelines provided in Sec. \ref{sec2}. Before going to, a very relevant point in the wormhole construction should be clarified. This argument is completely independent of the theory on which these solutions are studied. So, in principle the structure of a wormhole can be fixed for purely engineering reasons, without making any allusion to the type of matter that supports this object. This means that a throat connecting two infinite or asymptotically flat regions, can be obtained without solving the field equations as was done for example in \cite{Morris88}. Moreover, even if one is specifying the geometry of the wormhole and the general shape of the energy--momentum tensor of the matter distribution driven the configuration, it is not possible to guarantee that this matter content behaves as normal matter and respects the energy conditions \footnote{For a complete and recent revision of what energy conditions mean and its physical interpretation see \cite{curiel}.}.   

\subsection{The Class I Condition}

In general any theory of gravity contains a set of coupled non--linear partial differential equations as equations of motion. This
makes the task of solving them very difficult and therefore finding
exact solutions describing some situation of physical interest is really hard,
even if the spherical symmetry is imposed, which greatly reduces
the mathematical treatment. This inevitably requires the prescription of additional information to close the problem at least from the mathematical point of view. In this opportunity the system of equations given by Eqs. (\ref{eq9}) contains six unknowns, namely the geometry of the space--time $\{\Phi,b\}$, the thermodynamic variables $\{\rho, p_{r}, p_{\perp}\}$ and the $f(R)$ functional, hence it is necessary to fix three of them. In this case we shall fix the geometry $\{\Phi,b\}$ and the $f(R)$ gravity model. In order to fix $\{\Phi,b\}$ we shall employ the class I approach. As it is well--known, any spherically symmetric and static 4--dimensional manifold expressed by
\begin{equation}\label{eq18}
ds^{2}=-e^{\eta(r)}dt^{2}+e^{\lambda(r)}dr^{2}+r^2(d\theta^2+sin^2\theta d\phi^2)
\end{equation}
is of class II \cite{eisland,eisenhart}. This means that, it is necessary to have a 6--dimensional pseudo--Euclidean space to encrust it. However, under a suitable parametrization the space--time (\ref{eq18}) can be immersed into a 5--dimensional pseudo--Euclidean space, turning into a class I \cite{eisland,eisenhart,karmarkar}. In general any variety $V_{n}$ can be embedded in a flat space of $n\left(n+1\right)/2$ dimensions \cite{eisland,eisenhart}. Nevertheless, if the lowest order of this flat space is $n+p$, we say that $V_{n}$ is of class $p$. A theorem due to Eiesland \cite{eisland} establishes that a general centro--symmetric space--time
\begin{equation}\label{centro}
ds^{2}=-g_{1}dt^{2}+g_{2}dr^{2}+g_{3}(d\theta^2+sin^2\theta d\phi^2),    
\end{equation}
where $g_{i}=g_{i}(t,r)$ with $i=1,2,3$ are functions of the temporal $t$ and radial coordinate $r$, is of class I if and only if satisfies the following condition
\begin{equation}\label{eq244}
R_{t\,\theta\, t\,\theta}\,R_{r\,\phi\, r\,\phi}=R_{trtr}\,R_{\theta\,\phi\,\theta\,\phi}+R_{r\,\theta\,t\,\theta}R_{r\,\phi\,t\,\phi}. 
\end{equation}
In case of a static space--time, that is, $g_{i}=g_{i}(r)$, the line element becomes (\ref{centro})
\begin{equation}\label{centro1}
ds^{2}=-g_{1}dt^{2}+(1+g_{2})dr^{2}+r^2(d\theta^2+sin^2\theta d\phi^2).   
\end{equation}
Inserting (\ref{centro1}) into the condition (\ref{eq244}) one gets
\begin{equation}\label{class}
    g^{\prime\prime}_{1}=\frac{1}{2}\left[\frac{g^{\prime}_{1}g^{\prime}_{2}}{g_{2}}+\frac{\left(g^{\prime}_{1}\right)^{2}}{g_{1}}\right].
\end{equation}
The previous expression, corresponds to a first order differential equation in the potential $g_{2}$, but a non--linear second order differential equation in $g_{1}$. In any case, integration of this equation is possible, leading to $g_{1}=g_{1}(g_{2})$ or $g_{2}=g_{2}(g_{1})$. Specifically, one can obtain,
\begin{equation}\label{eqdif}
     \sqrt{g_{1}}=B+C\int \sqrt{g_{2}}dr,
\end{equation}
where $B$ and $C$ are integration constants. Comparing (\ref{eq18}) with (\ref{centro1}) and using (\ref{eqdif}) one gets\footnote{In the appendix \ref{A} it is provided a detailed derivation of the class I condition.}
\begin{equation}\label{classeta}
    e^{\eta(r)}=\left[B+C\int \sqrt{\left(e^{\lambda(r)}-1\right)}dr\right]^{2},
\end{equation}
or by inverting (\ref{eqdif}) one obtains
\begin{equation}\label{classlambda}
    e^{\lambda(r)}=1+A\eta^{\prime 2}(r)e^{\eta(r)},
\end{equation}
being $A$ an integration constant.
\subsection{The Model}

As it is observed, the class I condition provides a differential Equation (\ref{class}) linking both metric potentials, namely $e^{\eta}$ and $e^{\lambda}$. Once one of them is specified, the remaining one is directly obtained by means of (\ref{classeta}) or (\ref{classlambda}), obtaining in this way the full geometry of the desired space--time. Now, in comparing the line elements (\ref{eq1}) and (\ref{eq18}) it is clear that
\begin{equation}\label{eq30}
\Phi(r)=\eta(r), \quad b(r)=r\left(1-e^{-\lambda(r)}\right).    
\end{equation}
Therefore, in order to meet the wormhole throat condition i.e., $b(r_{0})=r_{0}$ from the right expression in (\ref{eq30}) it is evident that 
\begin{equation}\label{eq31}
    e^{-\lambda(r)}\bigg{|}_{r=r_{0}}=0.
\end{equation}
Furthermore, the asymptotically flat condition is satisfied if
\begin{equation}\label{eq32}
\lim_{r\rightarrow+\infty}e^{-\lambda(r)}\rightarrow 1 \Rightarrow \lim_{r\rightarrow+\infty}\frac{b(r)}{r}\rightarrow 0.
\end{equation}
However, to assure that the space--time behaves as the Minkowski space--time at large enough distances, also it is necessary to have
\begin{equation}\label{eq33}
\lim_{r\rightarrow+\infty}\Phi(r)=\Phi_{0},    
\end{equation}
being $\Phi_{0}$ a finite quantity. Then, after imposing the minimum requirements for a wormhole space--time, by replacing the expressions given by Equation (\ref{eq30}) into the Equations. (\ref{classeta})--(\ref{classlambda}) obtained from the class I condition (\ref{class}), one gets
\begin{equation}\label{eq34}
 b=r\left(1-\frac{1}{1+A\Phi^{\prime 2}e^{\Phi}}\right)   
\end{equation}
and
\begin{equation}\label{eq35}
e^{\Phi}=\left[B+C\int \sqrt{\frac{b}{r-b}}dr\right]^{2}.    
\end{equation}
To get the wormhole geometry, here we have adopted a suitable red--shift function $\Phi(r)$ given by \cite{raha}
\begin{equation}\label{eq36}
    \Phi(r)=\text{Ln}\bigg|\frac{\sqrt{\beta^{2}+r^{2}}}{r}\bigg|,
\end{equation}
being $\beta$ constant parameter with units of length.
Hence putting together (\ref{eq34}) and (\ref{eq36}) one gets the following shape function $b(r)$
\begin{equation}\label{eq37}
b(r)=\frac{Ar\beta^{4}}{r^{3}\sqrt{r^{2}+\beta^{2}}\left(r^{2}+\beta^{2}\right)+A\beta^{4}}.
\end{equation}
So, the wormhole space--time is given by
\begin{equation}\label{eq38}
ds^{2}=-\frac{\sqrt{\beta^{2}+r^{2}}}{r}dt^{2}+ \left(1-\frac{A\beta^{4}}{r^{3}\sqrt{r^{2}+\beta^{2}}\left(r^{2}+\beta^{2}\right)+A\beta^{4}}\right)^{-1}dr^{2}+  r^2(d\theta^2+sin^2\theta d\phi^2),
\end{equation}
where $A$ is a constant with units of $\text{length}^{2}$. At this point some comments are pertinent. First, the chosen red--shift function (\ref{eq36}) fulfills all the requirements listed in section \ref{sec2}. This function is finite everywhere for $r\in[r_{0},+\infty)$.
At the wormhole throat $r_{0}$ it is clear that $g_{tt}$ metric component of the line element (\ref{eq38}) is $e^{\Phi(r)}|_{r=r_{0}}\neq 0$, then the solution is free of event horizons. As $r\rightarrow+\infty$ then $\Phi(r)\rightarrow 0$, which implies that at large enough distances $g_{tt}=1$ when $r\rightarrow+\infty$. Second, concerning the form function $b(r)$, when the radial coordinate $r$ tends to infinite, the ratio $b(r)/r\rightarrow 0$. Therefore the manifold (\ref{eq38}) is asymptotically flat. Notwithstanding, the wormhole throat condition $b(r_{0})=r_{0}$ is not satisfied by (\ref{eq37}). To cure this problem, without loss of generality one can add a constant parameter\footnote{In general, this impasses in using the class I condition to get wormhole solutions, depends on the choice of the red--shift function $\Phi(r)$ \cite{K2}  (of course, this is the case when (\ref{classlambda}) is used to obtain the solution). For example, in \cite{tello} was proposed a suitable red--shift function satisfying all the conditions and leading to the correct shape function without including any extra parameter. So, in principle one cannot assure that the resulting  space--time meets the necessary and sufficient conditions to describe a wormhole. However, without loss of generality one can extend the solution to obtain the desired space--time \cite{K2}.} $\delta$. Then, (\ref{eq37}) becomes 
\begin{equation}\label{eq371}
b(r)=\frac{Ar\beta^{4}}{r^{3}\sqrt{r^{2}+\beta^{2}}\left(r^{2}+\beta^{2}\right)+A\beta^{4}}+\delta,
\end{equation}
where the solution of condition $b(r_{0})-r_{0}=0$ provides
\begin{equation}\label{delta}
\delta=\frac{r^{4}_{0}\left(r^{2}_{0}+\beta^{2}\right)^{3/2}}{r^{3}_{0}\sqrt{r^{2}_{0}+\beta^{2}}\left(r^{2}_{0}+\beta^{2}\right)+A\beta^{4}}.
\end{equation}
Therefore, the wormhole space--time is given by 
\begin{equation}\label{eq381}
ds^{2}=-\frac{\sqrt{\beta^{2}+r^{2}}}{r}dt^{2}+ \left(1-\frac{A\beta^{4}}{r^{3}\sqrt{r^{2}+\beta^{2}}\left(r^{2}+\beta^{2}\right)+A\beta^{4}}+\frac{r^{4}_{0}\left(r^{2}_{0}+\beta^{2}\right)^{3/2}}{rr^{3}_{0}\sqrt{r^{2}_{0}+\beta^{2}}\left(r^{2}_{0}+\beta^{2}\right)+Ar\beta^{4}}\right)^{-1}dr^{2}+  r^2(d\theta^2+sin^2\theta d\phi^2).
\end{equation}
\begin{figure}[tbph]
\centering
\includegraphics[width=0.35\textwidth]{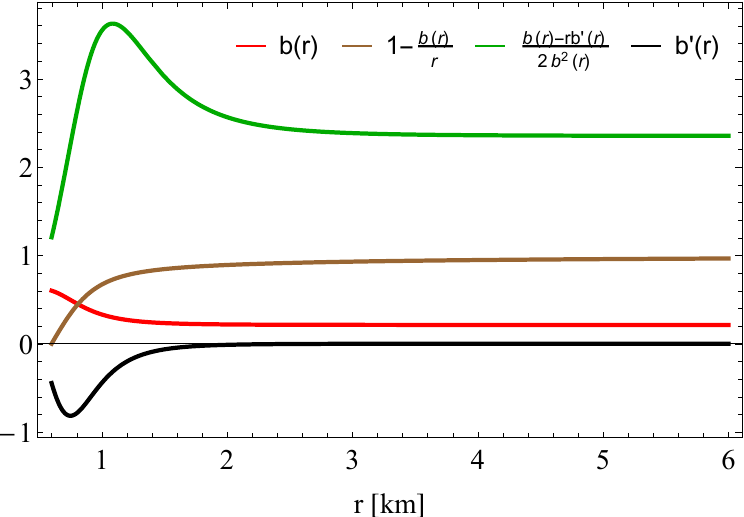} \ 
\includegraphics[width=0.37\textwidth]{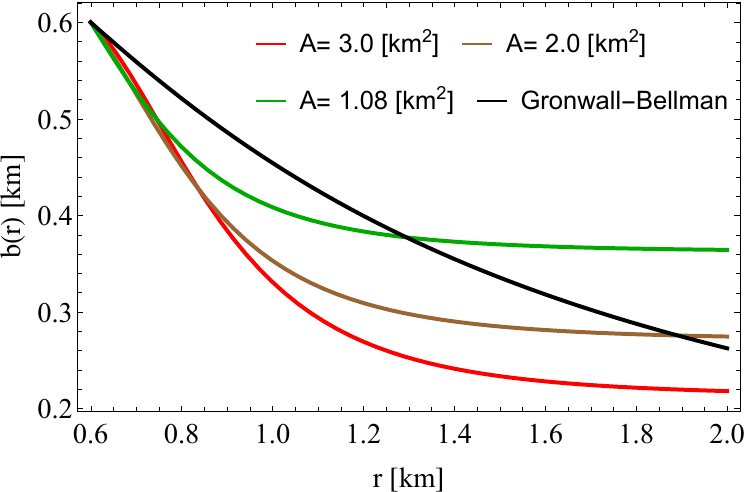}
\caption{\textbf{Left Panel}: The trend of the shape function (\ref{eq371}) versus the radial coordinate (red line). As can be seen, it is finite and regular for all $r\in[r_{0},+\infty)$. The brown line is depicting the asymptotically flat behavior of the $g^{-1}_{rr}$ metric component. The flare--out condition for the full range of the radial coordinate $r$ (green line) and at the wormhole throat (black line). In both cases, this condition is satisfied to ensure a traversable wormhole space--time. These plots were obtained by using $r_{0}=0.6$ [km], $\beta=0.5$ [km] and $A=3.0\, [\text{km}^{2}]$. It should be noted that for $b(r)$ the vertical axis has units of [km] whilst for the flare--out (green line) units of $[\text{km}^{-1}]$.
\textbf{Right Panel}: The trend of the shape function (\ref{eq371}) for different values of the constant $A$ and $\{r_{0};\beta\}=\{0.6;0.5\}\,[\text{km}]$. The black line corresponds to the form function obtained from the Gronwall--Bellman inequality. All the cases were obtained taking into account the same red--shift function (\ref{eq36}).  }
\label{fig1}
\end{figure}

As it is depicted in Figure \ref{fig1}, for $r_{0}=0.6$ [km], $\beta=0.5$ [km] and $A=3.0\, [\text{km}^{2}]$, the wormhole space--time (\ref{eq381}) is respecting all the features mentioned in section \ref{sec2}. It is observed that the shape function $b(r)$ (red curve) is starting from $0.6$ [km], the value of the wormhole throat $r_{0}$, tending to a finite quantity as the radial coordinate $r$ increases in magnitude. Taking into account the mentioned feature, brown line is exhibiting the asymptotically flat behavior of the solution, where at $r=r_{0}$ the quantity $1-b(r)/r$ is zero and for large enough distances tends to one. Another important characteristic in the construction of wormhole geometries, is the so--called flare--out condition \cite{Morris88,visser}. this condition is given by the following expression
\begin{equation}\label{eq39}
\frac{b(r)-rb^{\prime}(r)}{2b^{2}(r)}>0.
\end{equation}
Interestingly, at the wormhole throat the above expression provides $b^{\prime}(r_{0})\leq1$, where the equality holds only at the throat. So, inserting Equation (\ref{eq371}) into (\ref{eq39}) and evaluating at the wormhole throat $r_{0}$, one has
\begin{equation}\label{eq40}
\frac{A\beta^{4}\left(A\beta^{4}-5r^{5}_{0}\sqrt{\beta^{2}+r^{2}_{0}}-2r^{3}_{0}\sqrt{\beta^{2}+r^{2}_{0}}\right)}{\left(A\beta^{4}+5r^{5}_{0}\sqrt{\beta^{2}+r^{2}_{0}}+2r^{3}_{0}\sqrt{\beta^{2}+r^{2}_{0}}\right)^{2}}<1.   
\end{equation}
To ensure the satisfaction of the above expression, the constant $A$ must fulfill
\begin{equation}\label{flarecons}
    A>-\sqrt{\frac{r^{16}_{0}+5r^{14}_{0}\beta^{2}+10r^{12}_{0}\beta^{4}+10r^{10}_{0}\beta^{6}+5r^{8}_{0}\beta^{8}+r^{6}_{0}\beta^{10}}{\beta^{8}\left(7r^{2}_{0}+4\beta^{2}\right)^{2}}}.
\end{equation}
For the chosen values, the expression (\ref{eq39}) is valid everywhere (see green line in Figure \ref{fig1}), whilst the same condition at the wormhole throat (\ref{eq40}) (black line in Figure \ref{fig1}) is also satisfied. It should be pointed out, that the flare--out condition is a necessary condition to have a traversable wormhole solution, however, it is not enough to ensure a humanely traversable wormhole space--time (see below) \cite{Morris88,visser} .

\begin{figure}[H]
\centering
\includegraphics[width=0.35\textwidth]{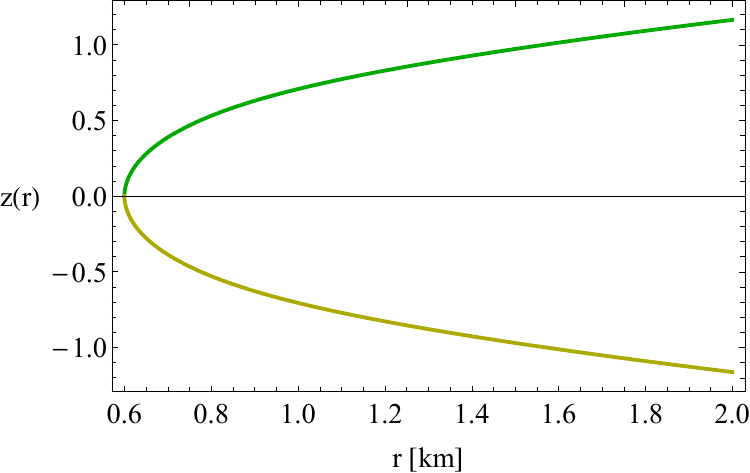} 
\
\includegraphics[width=0.42\textwidth]{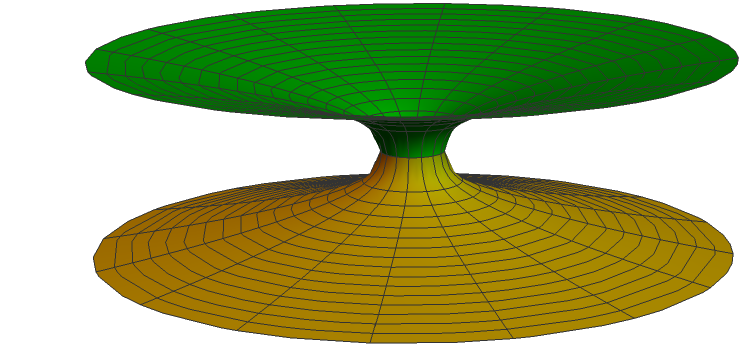}
\caption{\textbf{Left panel}: The 2D embedding diagram, the $z(r)$ function versus the radial coordinate $r$. \textbf{Right panel}: The 3D wormhole geometry, obtained by rotating the 2D $z(r)$ function around a vertical axis. These 2D and 3D embedding diagrams were plotted by considering the numerical values $r_{0}=0.6$ [km], $\beta=0.5$ [km] and $A=3.0\, [\text{km}^{2}]$. }
\label{fig3}
\end{figure}

Finally, to close the geometrical description we use the embedding diagrams to represent the wormhole space--time and extract some
information about the obtained shape function $b(r)$ given by (\ref{eq371}). Since we are dealing with spherically symmetric and static wormhole solutions, from the general line element (\ref{eq1}) focusing on an equatorial plane, $\theta=\pi/2$, the solid angle element $d\Omega^{2}\equiv d\theta^2+sin^2\theta d\phi^2$ reduces to
\begin{equation}\label{eq41}
d\Omega^{2}=d\phi^{2}.
\end{equation}
Besides, by fixing a constant time slide i.e., $t=\text{constant}$, the line element (\ref{eq1}) becomes
\begin{equation}\label{eq42}
ds^{2}=\frac{dr^{2}}{1-\frac{b(r)}{r}}+r^{2}d\phi^{2}.    
\end{equation}
To visualize this equatorial plane as a surface embedded in an Euclidean space, it is convenient to introduce cylindrical coordinates as \cite{Morris88,visser}
\begin{equation}\label{eq43}
ds^{2}=dz^{2}+dr^{2}+r^{2}d\phi^{2},    
\end{equation}
or, equivalently,
\begin{equation}\label{eq44}
ds^{2}=\left[1+\left(\frac{dz}{dr}\right)^{2}\right]dr^{2}+r^{2}d\phi^{2}.    
\end{equation}
Now, comparing (\ref{eq42}) and (\ref{eq44}), one obtains
\begin{equation}\label{eq45}
\frac{dz}{dr}=\pm\left(\frac{r}{b(r)}-1\right)^{-1/2},    
\end{equation}
where the function $z=z(r)$ defines the embedded surface \cite{Morris88,visser}. Interestingly, when $r\rightarrow+\infty$ the expression (\ref{eq45}) leads to
\begin{equation}\label{eq46}
    \frac{dz}{dr}\bigg|_{r\rightarrow+\infty}=0,
\end{equation}
which tells us that the embedding diagram provides two asymptotically flat patches. Of course, this is the case as was pointed out before. In addition, note that the integration of Equation (\ref{eq45}), using (\ref{eq371}), can not be done analytically. Therefore, a numerical treatment has been done in order to illustrate the wormhole shape given in Figure \ref{fig3}. The left panel of Figure \ref{fig3} is showing the 2D embedding diagram, the $z(r)$ function. The green line is representing the upper universe or spatial station and the yellow one the lower universe or spatial station. The rotation of this
curve around a vertical axis, generates the wormhole hyper–-surface (Figure \ref{fig3} right panel). As can be seen, the form of this hyperboloid is coherent with the previous discussion, exhibiting a tunnel connecting two asymptotically flat spaces, hence the present model has the usual or trivial wormhole topology  $\mathcal{M}\sim\mathbb{R}\times\mathbb{R}\times \mathbb{S}^{2}$, that is, the time $\mathbb{R}$, the radius $\mathbb{R}$ and the throat with a two sphere $\mathbb{S}^{2}$ topology. This kind of object also are called Lorentzian space--times with a compact region $\mathcal{M}$ containing a quasipermanent intra--universe wormhole \cite{visser}. Besides, in this case the two connected regions by means of the tunnel have a $\mathbb{R}^{3}$ trivial topology. In this concern, in principle these regions might not be a Minkowski space--time. In fact, as was shown in \cite{roman}, these spaces can be a de Sitter space in case of dynamical wormhole structures.

To close the geometrical description of the model, some comments are in order. First, by using merely engineering arguments in conjunction with the class I condition (\ref{classlambda}), it is possible to build up well--established wormhole structures. Of course, as was mentioned earlier one cannot assure that the matter content driven this space--time is respecting the so--called energy conditions. However, it is evident that the class I approach helps to reduce the number of unknowns under suitable and well motivated choices of the metric components, $e^{\eta}$ and $e^{\lambda}$, to close the problem at least mathematically.

\subsection{The Energy Conditions and Ghost Fields}\label{sec5}

\begin{figure}[H]
\centering
\includegraphics[width=0.35\textwidth]{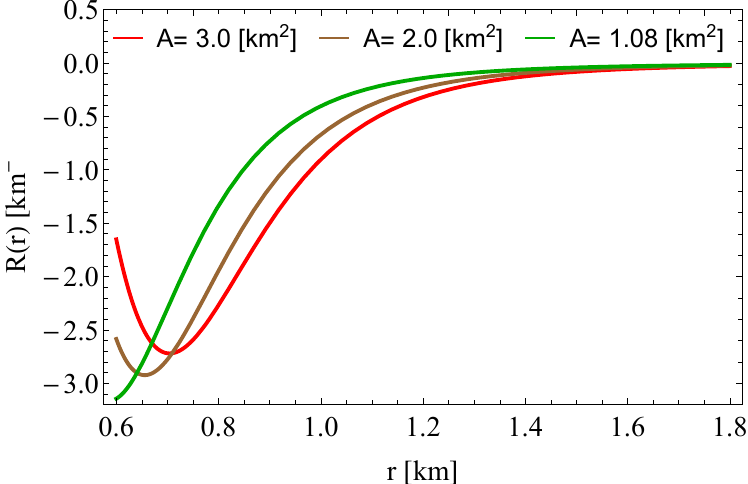} \ 
\includegraphics[width=0.32\textwidth]{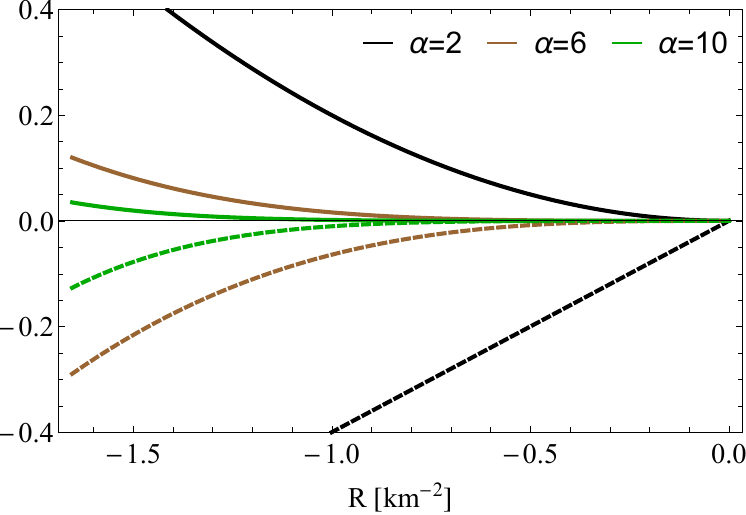}
\caption{\textbf{Left panel}: The trend of the Ricci scalar $R$ versus the radial coordinate $r$ for different values of the parameter $A$ and $\{r_{0};\beta\}=\{0.6;0.5\}\,[\text{km}]$. \textbf{Right panel}: The $f(R)$ gravity model (solid lines) and its first derivative $F(R)$ (dashed lines) against the scalar curvature $R$, for different values of the constant $\alpha$. To obtain these plots $A=3.0\,[\text{km}^{2}]$ and $\{r_{0};\beta\}=\{0.6;0.5\}\,[\text{km}]$.}
\label{fig5}
\end{figure}

\begin{figure}[H]
\centering
\includegraphics[width=0.35\textwidth]{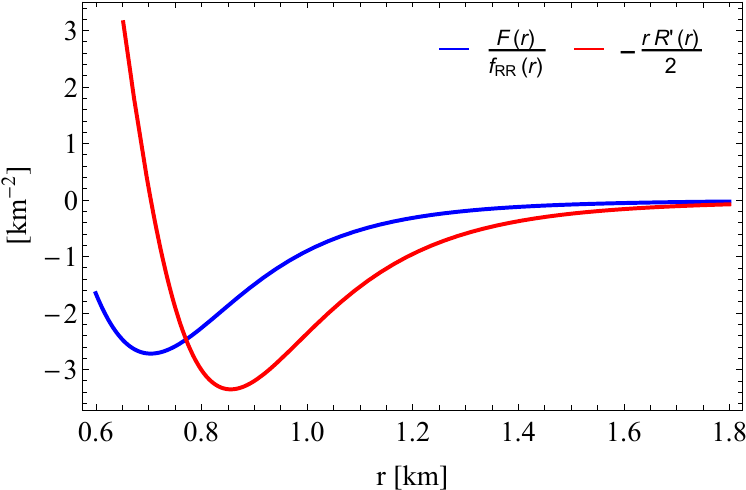} \ 
\includegraphics[width=0.35\textwidth]{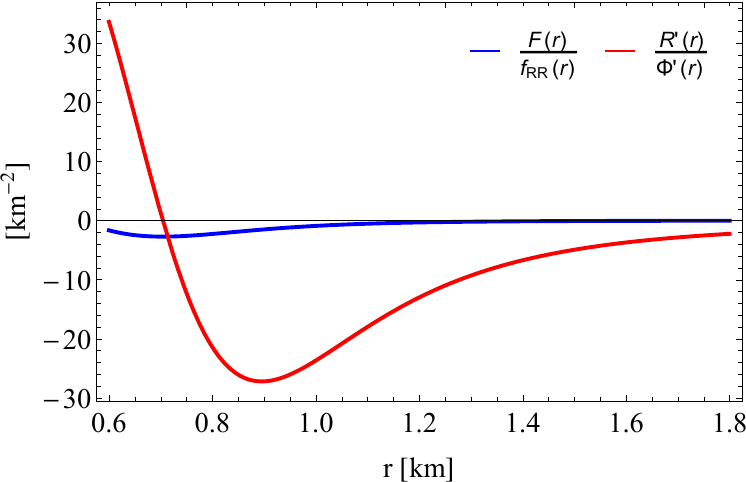}
\caption{\textbf{Left panel}: The condition (\ref{conditionwecr}) versus the radial coordinate $r$.  \textbf{Right panel}: The constraint (\ref{cond1}) against the radial coordinate $r$. These plots were obtained by considering $A=3.0\,[\text{km}^{2}]$, $\chi=2.5\,[\text{km}^{-2}]$, $\{r_{0};\beta\}=\{0.6;0.5\}\,[\text{km}]$ and $\alpha=2.0$.   }
\label{fig6}
\end{figure}

As was pointed out above, the fact of having a wormhole geometry driven by a matter distribution satisfying energy conditions everywhere, is extremely difficult if not impossible. Moreover, from Equations (\ref{eq9}) it is not easy to establish general constraints on the thermodynamic variables $\rho$, $p_{r}$ and $p_{\perp}$, in order to fulfill NEC and WEC. However, by evaluating these expressions at the wormhole throat $r=r_{0}$,
one can get valuable information for some constant parameters characterizing the model and, in this way get the satisfaction of NEC and WEC at $r=r_{0}$ and its neighborhood. So, from Equations (\ref{eq9}) one obtains the following reduced expressions for $\rho$, $\rho+p_{r}$ and $\rho+p_{\perp}$ at $r=r_{0}$, where $b(r)|_{r=r_{0}}=r_{0}$, \cite{Pavlovic15}
\begin{equation}\label{eq471}
\begin{split}
\rho(r)=\frac{1}{2}f(r)+\frac{rb^{\prime}(r)-r}{2r^{2}}\left[f_{RR}(r)R^{\prime}(r)-F(r)\Phi^{\prime}(r)\right]\geq 0,
\end{split}
\end{equation}
\begin{eqnarray}\label{eq47}
\rho(r)+p_{r}(r)&=&\frac{rb^{\prime}(r) -r}{2 r^{3}}\bigg[f_{R R}(r) R^{\prime}(r)r +2F(r)\bigg]\geq 0, \\ \label{eq48}
\rho(r)+p_{\perp}(r)&=&\frac{F(r)}{r^{2}}\left[ \frac{rb^{\prime}(r) +r}{2 r}-\frac{\Phi^{\prime}(r)\left( rb^{\prime}(r) -r\right)}{2 }\right]\geq 0.
\end{eqnarray}
As can be appreciated, on the right member of Equation (\ref{eq47}) appears the flare--out condition (\ref{eq39}). Given that this condition is always positive, thus $ (rb^{\prime}(r)-r)|_{r=r_{0}}$ is negative. So, to guarantee a positive $\rho+p_{r}$ at the wormhole throat and its neighborhood one needs to assure 
\begin{equation}\label{conditionwecr}
    f_{R R}(r) R^{\prime}(r) r+2F(r)\leq0 \Rightarrow \frac{F(r)}{f_{RR}(r)}\leq -\frac{rR^{\prime}(r)}{2}.
\end{equation}
It is obvious that this
condition strongly depends on the choice of $f(R)$ gravity model and, cannot be satisfied for every $f(R)$ functional. As we know if $f(R)=R$, Einstein theory, the expression (\ref{eq47}) is automatically transgressed.
Besides, as we are interested in regions where $F<0$ (violation of the non--existence theorem \cite{Bronnikov07,Bro}), the bracket of the second member in Equation (\ref{eq48}) provides an interesting constraint on the shape function $b(r)$ near the throat, namely
\begin{equation}\label{gronwallinequality}
  b^{\prime}(r)\leq \frac{b(r)\left(1+r\Phi^{\prime}(r)\right)}{r^{2}\Phi^{\prime}(r)-r}.
\end{equation}
As $\Phi(r)$ is a real valued, continuous and differentiable function on the interval $[r_{0},+\infty)$, then the function $\frac{1+r\Phi^{\prime}(r)}{r^{2}\Phi^{\prime}(r)-r}$ is too. Furthermore, as $b(r)$ is differentiable in the interior of the mentioned interval, that is, for all $r\in (r_{0},+\infty)$. From the inequality (\ref{gronwallinequality}) one gets
\begin{equation}\label{groned}
    b(r)\leq b(r_{0})\text{Exp}\left(\int^{r}_{r_{0}}\frac{1+x\Phi^{\prime}(x)}{x^{2}\Phi^{\prime}(x)-x}dx\right).
\end{equation}
The above result comes from the so--called Gronwall--Bellman (GB) inequality or Gronwall's lemma \cite{Pavlovic15}. So, it is clear that the red--shift function $\Phi(r)$ play a major role in the satisfaction of both the NEC and WEC. So, in general the problem of having traversable wormhole solutions in the $f(R)$ gravity scenario satisfying at least the NEC and WEC, upon depends on several constraints imposed on the $f(R)$ model, flare--out condition, shape function and red--shift, to name a few. Therefore, the main point behind this analysis, is to restrict the pair $\{b(r),\Phi(r)\}$ in order to satisfy (\ref{eq471})--(\ref{eq48}) at the same footing of the requirements listed in Sec. \ref{sec2}. So, we start analyzing the differential inequality (\ref{groned}). Taking the critical case i.e., the equality symbol and plugging the red--shift function (\ref{eq36}) one gets
\begin{equation}\label{shapegron}
b(r)_{\text{GB}}=r_{0}\sqrt{\frac{r^{2}_{0}+2\beta^{2}}{r^{2}+2\beta^{2}}}.    
\end{equation}
Of course, the solution (\ref{shapegron}) is a particular solution of the differential inequality (\ref{groned}). Notwithstanding, it is sufficient to establish whether the shape function (\ref{eq371}) obtained from the class I methodology is leading or not to the satisfaction of the Equation (\ref{eq48}). This is possible if and only if (\ref{eq371}) in bounded from above by (\ref{shapegron}) at the wormhole throat and its neighborhood. The left panel of Figure \ref{fig1}, corroborates that $b(r)\leq b(r)_{GB}$ for the chosen red--shift (\ref{eq36}) and different values of the parameter $A$. Additionally, to fulfill (\ref{eq48}) one needs the full information about the $f(R)$ gravity model. In this regard, we have selected the following model \cite{Tripathy16} 
\begin{equation}\label{eq49}
f(R)=\left(\frac{R}{\chi}\right)^{\alpha/2}\left(\frac{2R}{2+\alpha}\right),
\end{equation}
where $\alpha$ (dimensionless) and $\chi$ (with units of $[\text{km}^{-2}]$) are constant free parameters and, $R$ is the Ricci scalar curvature. This model was proposed to describe anisotropic cosmological $f(R)$ gravity models, considering plane symmetric models with anisotropy in the expansion rates. The anisotropy in expansion rates were assumed to be maintained throughout the cosmic evolution. Now, the function $F(R)$ is given by
\begin{equation}\label{F(R)}
    F(R)=\left(\frac{R}{\chi}\right)^{\alpha/2}.
\end{equation}
It is evident that the condition $F(R)<0$ (violation of the non--existence theorem), occurs when $R<0$ and $\chi>0$ (or vice--versa) subject to $\alpha/2 \in \mathbb{Z}$ and $\alpha=4n+2$, with $n\in \mathbb{N}_{0}$, to avoid imaginary values. For the particular value $\alpha=2.0$ in conjunction with the case $\chi>0$, then automatically $R<0$, Equation (\ref{F(R)}) provides the following bounds on the constant $A$ in terms of $r_{0}$ and $\beta$\footnote{Actually, to constraint $A$ in terms of $r_{0}$ and $\beta$, one should use the scalar curvature $R$ instead of $F(R)$. This is so because, $F(R)$ depends on $A$, $r_{0}$ and $\beta$ through $R$. Then, (\ref{F(R)}) is equivalent to impose $R<0$ or $R>0$, depending whether $\chi>0$ or $\chi<0$, respectively.},
\begin{equation}\label{Fcondition}
\begin{split}
    \bigg(\frac{200 r_{0}^{14} \beta ^4}{16 r_{0}^2 \beta ^{12}+16
\beta ^{14}}+\frac{416 r_{0}^{12} \beta ^6}{16 r_{0}^2 \beta ^{12}+16 \beta ^{14}}+\frac{577 r_{0}^{10} \beta ^8}{2 \left(16 r_{0}^2
\beta ^{12}+16 \beta ^{14}\right)}+\frac{72 r_{0}^8 \beta ^{10}}{16 r_{0}^2 \beta ^{12}+16 \beta ^{14}}+\frac{4 r_{0}^6 \beta ^{12}}{16
r_{0}^2 \beta ^{12}+16 \beta ^{14}}&\\-\frac{10 r_{0}^{11} \beta ^4 \sqrt{400 r_{0}^6+824 r_{0}^4 \beta ^2+553 r_{0}^2 \beta ^4+120
\beta ^6}}{16 r_{0}^2 \beta ^{12}+16 \beta ^{14}}-\frac{21 r_{0}^9 \beta ^6 \sqrt{400 r_{0}^6+824 r_{0}^4 \beta ^2+553 r_{0}^2
\beta ^4+120 \beta ^6}}{2 \left(16 r_{0}^2 \beta ^{12}+16 \beta ^{14}\right)}&\\-\frac{2 r_{0}^7 \beta ^8 \sqrt{400 r_{0}^6+824 r_{0}^4
\beta ^2+553 r_{0}^2 \beta ^4+120 \beta ^6}}{16 r_{0}^2 \beta ^{12}+16 \beta ^{14}}\bigg)^{1/2}< A<\bigg(\frac{200 r_{0}^{14} \beta ^4}{16 r_{0}^2 \beta ^{12}+16 \beta ^{14}}+\frac{416 r_{0}^{12} \beta ^6}{16 r_{0}^2 \beta
^{12}+16 \beta ^{14}}&\\+\frac{10 r_{0}^{11} \beta
^4 \sqrt{400 r_{0}^6+824 r_{0}^4 \beta ^2+553 r_{0}^2 \beta ^4+120 \beta ^6}}{16 r_{0}^2 \beta ^{12}+16 \beta ^{14}}+\frac{21 r_{0}^9
\beta ^6 \sqrt{400 r_{0}^6+824 r_{0}^4 \beta ^2+553 r_{0}^2 \beta ^4+120 \beta ^6}}{2 \left(16 r_{0}^2 \beta ^{12}+16 \beta ^{14}\right)}&\\+\frac{2
r_{0}^7 \beta ^8 \sqrt{400 r_{0}^6+824 r_{0}^4 \beta ^2+553 r_{0}^2 \beta ^4+120 \beta ^6}}{16 r_{0}^2 \beta ^{12}+16 \beta ^{14}}+\frac{577 r_{0}^{10} \beta ^8}{2 \left(16 r_{0}^2 \beta ^{12}+16 \beta ^{14}\right)}+\frac{72 r_{0}^8 \beta ^{10}}{16
r_{0}^2 \beta ^{12}+16 \beta ^{14}}&\\+\frac{4 r_{0}^6 \beta ^{12}}{16 r_{0}^2 \beta ^{12}+16 \beta ^{14}}\bigg)^{1/2}.
\end{split}
\end{equation}
As can be seen, the above constraints on $A$ are independent of the parameter $\chi$, but its signature is relevant to get the desired condition $F(R)<0$. As mentioned, if $\chi>0$ the scalar curvature $R$ must be negative. Indeed, as the left panel of Figure \ref{fig5}
shows, the Ricci scalar is negative, starting from the wormhole throat and beyond it, reaching its minimum value at infinite. Specifically, when $r\rightarrow+\infty$ then $R(r)\rightarrow 0$. On the other hand, the right panel of Figure \ref{fig5} depicts the behavior of the $f(R)$ (solid lines) gravity model and its first derivative $F(R)$ (dashed lines). As it is observed, from the throat towards large enough distances, the $f(R)$ is positive and decreasing function with increasing scalar curvature. This agree with the negative nature of $F(R)$. Then, at this point one can ensure the fulfillment of the expression (\ref{eq48}).
Now, the expression (\ref{eq47}) entails an interesting situation. As said before, the global term on the right hand side involves the flare--out condition at throat. As $b'(r_{0})<1$, then this global factor is negative in nature (see black line in Figure \ref{fig1}). Therefore, we need to check the condition (\ref{conditionwecr}). From the left panel of Figure \ref{fig5} it is clear that the Ricci scalar is decreasing at the wormhole throat and its vicinity, hence its derivative $R'(r)$ is negative, in this way the right member of (\ref{conditionwecr}) is positive. Moreover, as $F(R)$ is negative and increasing function (see right panel of Figure \ref{fig5}), then $f_{RR}$ is positive. So, the left member of (\ref{conditionwecr}) is negative. Thus, for the present model one can assure that the expression (\ref{eq47}) is satisfied. This information can be corroborated in the left panel of Figure \ref{fig6} where it is clear that $F/f_{RR}\leq -rR'/2$ at the throat and its neighborhood. Finally, for the density given by Equation (\ref{eq471}), one needs to analyze every member of the right hand side and then glue the result for each term in order to fix the proper conditions to have a positive density $\rho(r)$. The simplest case is to consider the $f(R)$ model to be positive defined (depending on the form of the $f(R)$ model one can always force it to be positive). In doing this and taking into account the previous assumptions, from Equation (\ref{eq49}) it is clear that $f(R)>0$, if $\alpha>-2$. Nevertheless, as $\alpha/2\in \mathbb{Z}$ then $\alpha$ cannot be $-1$, also the case $\alpha=0$ yields to GR, then energy conditions are violated, hence these values are ruled out. So, the only possibility to guarantee $f(R)>0$ is to consider $\alpha/2\in \mathbb{Z}^{+}$. Next, the second member of the right hand side of expression (\ref{eq471}) again contains the flare--out condition, then the bracket should be negative. To warrant this requirement, it is necessary to analyze the behavior of the red--shift function. As we are looking from the very beginning for asymptotically flat and traversable wormholes\footnote{In case of having a non--asymptotically flat wormhole, this issue can be solved by cutting and pasting the resulting space--time with an other one, satisfying the desired requirements.}, the red--shift function should be finite or zero when the radial coordinate tends to infinite. This condition can be reached if $\Phi(r)$ is positive (in order to avoid event horizons) and decreasing in nature, in such a case $\Phi'(r)<0$. Although, if $\Phi(r)$ is increasing it is possible to get a finite value at large enough distances, however, in that case the time coordinate should be redefined to mimic the Minkowski space--time. In the present case, the red--shift function (\ref{eq36}) is decreasing in nature. This can be easily checked by computing its first derivative. So,
\begin{equation}
\Phi'(r)=-\frac{\beta^{2}}{r\left(\beta^{2}+r^{2}\right)}.
\end{equation}
So, as $\Phi^{\prime}(r)<0$, hence from Equation (\ref{eq471}) one gets the following restriction
\begin{equation}\label{cond1}
    \frac{F(r)}{f_{RR}(r)}<\frac{R'(r)}{\Phi'(r)}.
\end{equation}
Hence, the right member of the inequality (\ref{cond1}) is positive and consequently greater that the ratio $F/f_{RR}$. This can be observed in the right panel of Figure \ref{fig6}, where $R'/\Phi'$ overcomes $F/f_{RR}$ at the wormhole throat and its vicinity. At this point it is worth mentioning that, we have considered the simplest scenario to analyze the positiveness of the density $\rho(r)$, however it is evident that all these things strongly depend on the geometry of the space--time and the $f(R)$ gravity model.  

Now, the above assumptions in conjunction with expressions (\ref{eq471})--(\ref{eq48}) lead to the following bounds on the constant $A$:\\

$\rho\geq0$,\\
\begin{equation}\label{dencondition}
    A\geq \mathbb{Q}^{(10)}(x=7;r_{0};\beta),
\end{equation}
$\rho+p_{r}\geq 0$,\\

\begin{equation}\label{wecrcondition}
A\geq \mathbb{P}^{(4)}(x=4;r_{0};\beta).
\end{equation}
The above bounds have been written in compact form since the expressions are extremely large to be displayed here. The notation $\mathbb{Q}^{(10)}$ and $\mathbb{P}^{(4)}$ correspond to polynomial expressions of degree ten and four in the variable $x$ respectively, depending on the parameters $r_{0}$ and $\beta$. For the former, the bound corresponds to the seventh root of the polynomial whilst in the second case, the bound corresponds to the fourth root of $\mathbb{P}^{(4)}$. Of course, in all cases only real roots are considered. \\

\begin{figure}[H]
\centering
\includegraphics[width=0.32\textwidth]{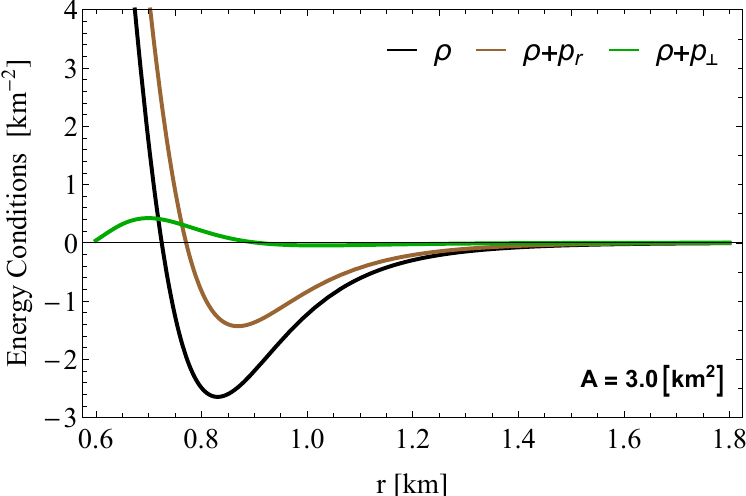} \ 
\includegraphics[width=0.32\textwidth]{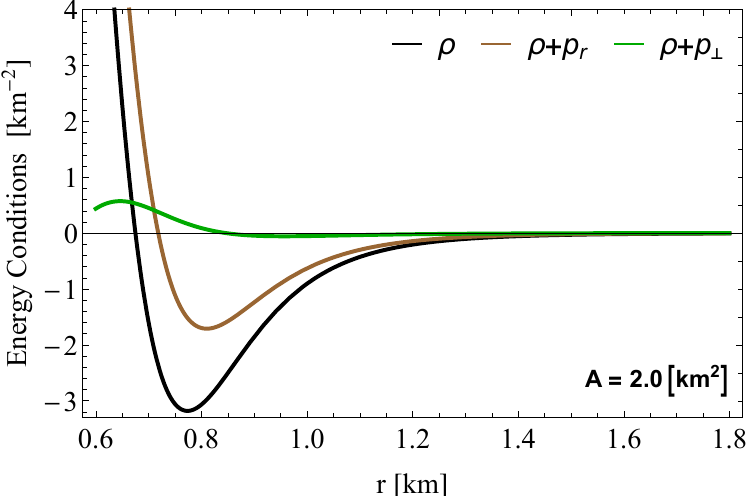}\
\includegraphics[width=0.32\textwidth]{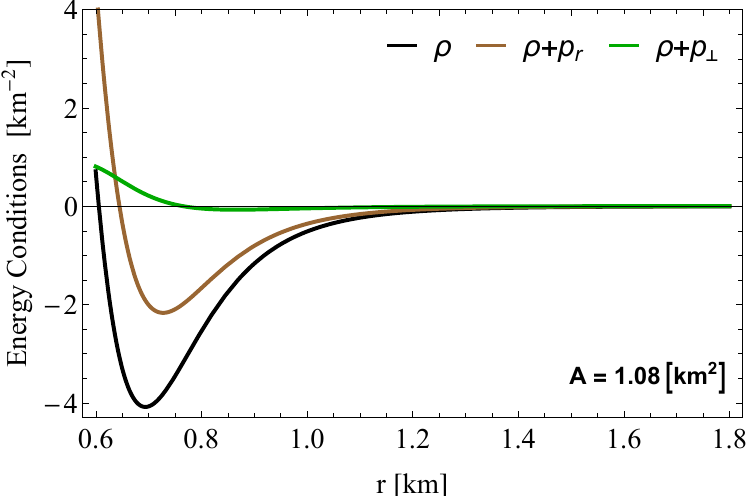} 
\caption{The null and weak energy conditions versus the radial coordinate $r$, for different values of the parameter $A$ and considering $\chi=2.5\,[\text{km}^{-2}]$, $\{r_{0};\beta\}=\{0.6;0.5\}\,[\text{km}]$ and $\alpha=2.0$.  }
\label{fig7}
\end{figure}

\begin{figure}[H]
\centering
\includegraphics[width=0.32\textwidth]{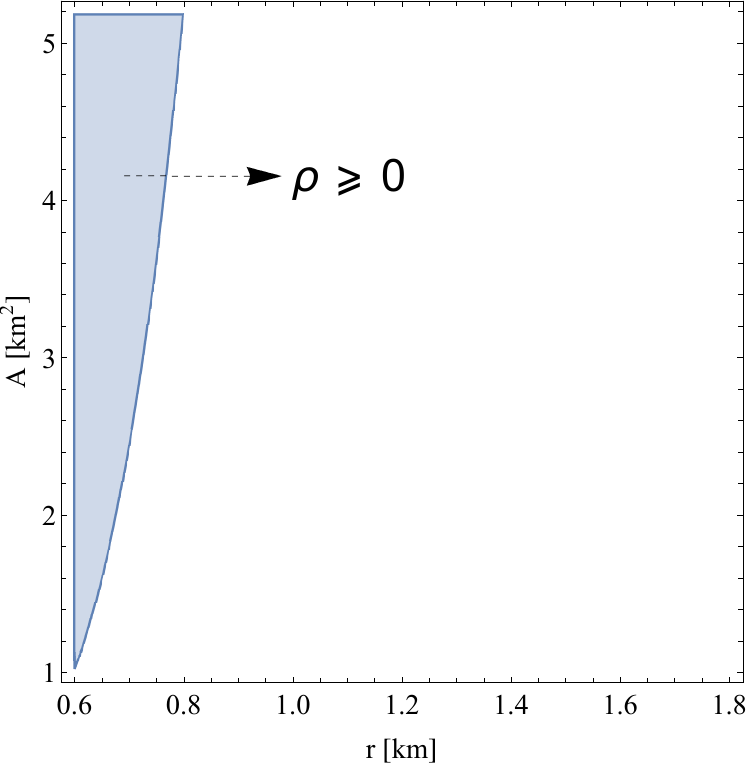} \ 
\includegraphics[width=0.32\textwidth]{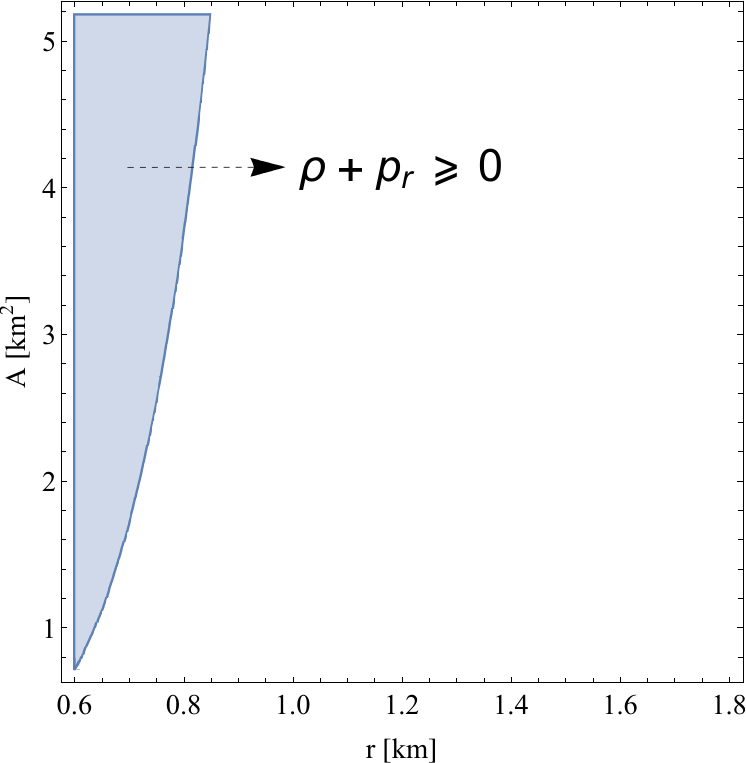}\
\includegraphics[width=0.33\textwidth]{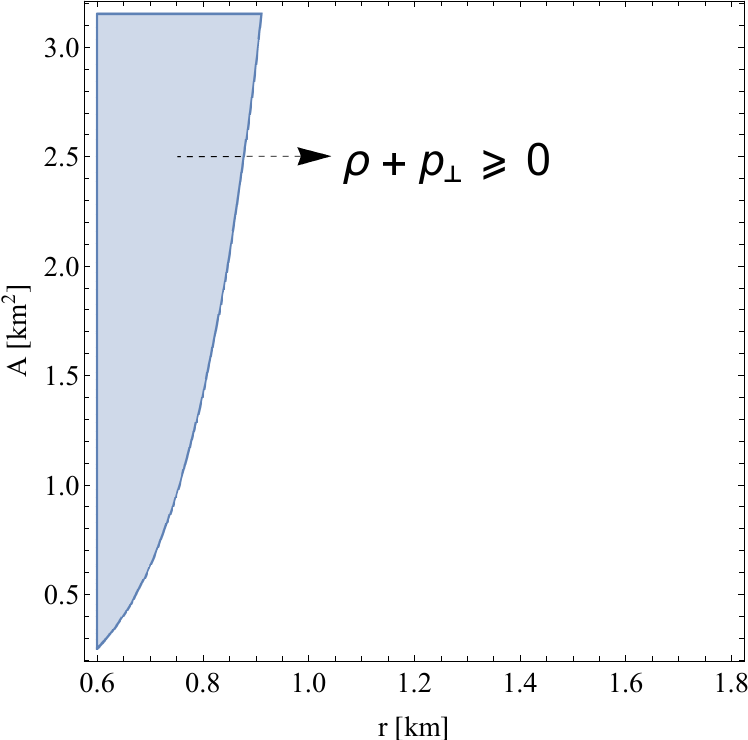} 
\caption{\textbf{Left panel}: Region plots for $\rho\geq 0$ for $A\geq 1.024$. The blue region shows the satisfaction $\rho\geq 0$ at the throat for our proposed toy model. \textbf{Middle panel} and \textbf{Right panel}: Blue regions are displaying the satisfaction of $\rho+p_{r}\geq 0$ and $\rho+p_{\perp}\geq 0$ at the wormhole throat. In the first case $A\geq 0.718$ and in the second one $0.254\leq A\leq 3.151$. It should be taken into account that for the left and middle panels, the upper bound for $A$ is imposed by the condition $F <0$ \i.e., $A\geq 5.181$. These region plots were obtained by considering $\chi=2.5\,[\text{km}^{-2}]$, $\{r_{0};\beta\}=\{0.6;0.5\}\,[\text{km}]$ and $\alpha=2.0$.}
\label{fig8}
\end{figure}

\begin{figure}[H]
\centering
\includegraphics[width=0.35\textwidth]{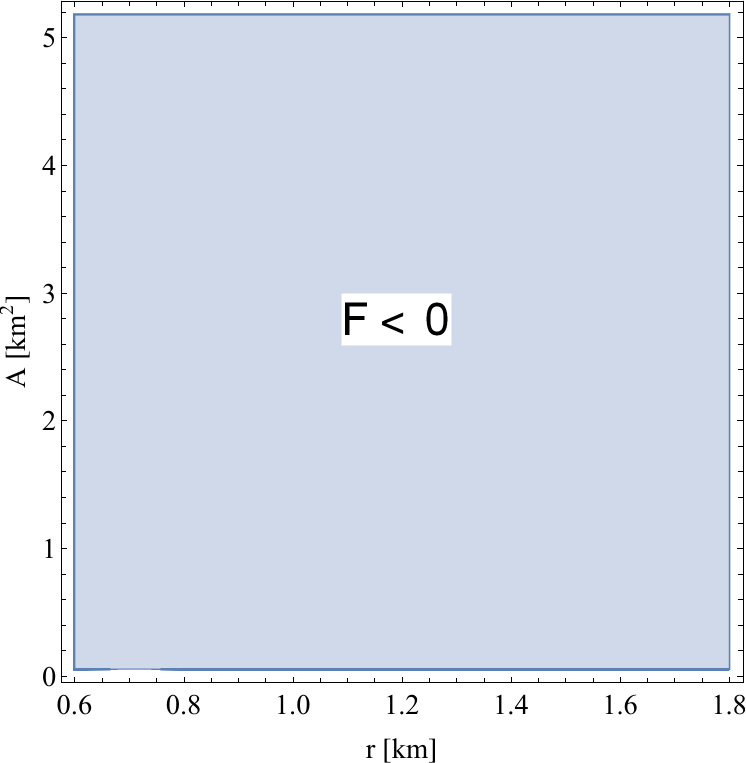} \ 
\includegraphics[width=0.35\textwidth]{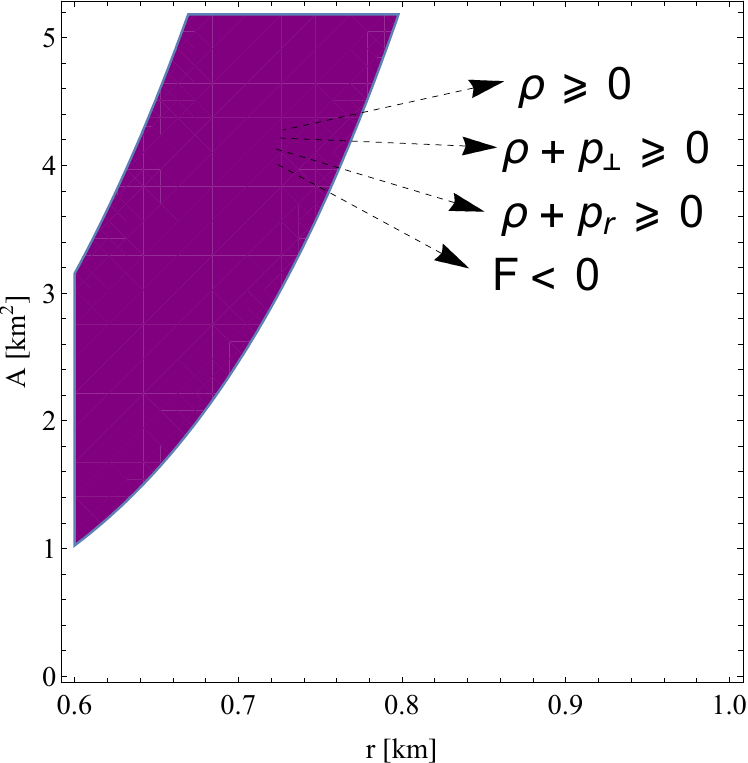}
\caption{\textbf{Left panel}: Region plot for $F<0$, taking into account the $f(R)$ gravity model (\ref{eq49}) and $0.054\leq A\leq 5.181$. \textbf{Right panel}: The common region for the satisfaction at the same footing of NEC, WEC and the non--existence theorem violation, for the present toy model, considering $0.054\leq A\leq 5.181$ and $\chi=2.5\,[\text{km}^{-2}]$, $\{r_{0};\beta\}=\{0.6;0.5\}\,[\text{km}]$ and $\alpha=2.0$.}
\label{fig9}
\end{figure}

$\rho+p_{\perp}\geq 0$,

\begin{equation}\label{wectcondition}
    \begin{split}
        \bigg(\frac{5 r_{0}^{14} \beta
^8}{2 \left(4 r_{0}^2 \beta ^{16}+4 \beta ^{18}\right)}+\frac{24 r_{0}^{12} \beta ^{10}}{4 r_{0}^2 \beta ^{16}+4 \beta ^{18}}+\frac{56
r_{0}^{10} \beta ^{12}}{4 r_{0}^2 \beta ^{16}+4 \beta ^{18}}+\frac{38 r_{0}^8 \beta ^{14}}{4 r_{0}^2 \beta ^{16}+4 \beta ^{18}}+\frac{8
r_{0}^6 \beta ^{16}}{4 r_{0}^2 \beta ^{16}+4 \beta ^{18}}&\\-\frac{3 r_{0}^{10} \beta ^8 \sqrt{r_{0}^8+36 r_{0}^6 \beta ^2+100 r_{0}^4
\beta ^4+72 r_{0}^2 \beta ^6+16 \beta ^8}}{2 \left(4 r_{0}^2 \beta ^{16}+4 \beta ^{18}\right)}-\frac{5 r_{0}^8 \beta ^{10} \sqrt{r_{0}^8+36
r_{0}^6 \beta ^2+100 r_{0}^4 \beta ^4+72 r_{0}^2 \beta ^6+16 \beta ^8}}{4 r_{0}^2 \beta ^{16}+4 \beta ^{18}}&\\-\frac{2 r_{0}^6
\beta ^{12} \sqrt{r_{0}^8+36 r_{0}^6 \beta ^2+100 r_{0}^4 \beta ^4+72 r_{0}^2 \beta ^6+16 \beta ^8}}{4 r_{0}^2 \beta ^{16}+4
\beta ^{18}}\bigg)^{1/2}\leq A\leq \bigg(\frac{5 r_{0}^{14}
\beta ^8}{2 \left(4 r_{0}^2 \beta ^{16}+4 \beta ^{18}\right)}+\frac{24 r_{0}^{12} \beta ^{10}}{4 r_{0}^2 \beta ^{16}+4 \beta ^{18}}&\\+\frac{56
r_{0}^{10} \beta ^{12}}{4 r_{0}^2 \beta ^{16}+4 \beta ^{18}}+\frac{38 r_{0}^8 \beta ^{14}}{4 r_{0}^2 \beta ^{16}+4 \beta ^{18}}+\frac{8
r_{0}^6 \beta ^{16}}{4 r_{0}^2 \beta ^{16}+4 \beta ^{18}}+\frac{3 r_{0}^{10} \beta ^8 \sqrt{r_{0}^8+36 r_{0}^6 \beta ^2+100 r_{0}^4
\beta ^4+72 r_{0}^2 \beta ^6+16 \beta ^8}}{2 \left(4 r_{0}^2 \beta ^{16}+4 \beta ^{18}\right)}&\\+\frac{5 r_{0}^8 \beta ^{10} \sqrt{r_{0}^8+36
r_{0}^6 \beta ^2+100 r_{0}^4 \beta ^4+72 r_{0}^2 \beta ^6+16 \beta ^8}}{4 r_{0}^2 \beta ^{16}+4 \beta ^{18}}+\frac{2 r_{0}^6
\beta ^{12} \sqrt{r_{0}^8+36 r_{0}^6 \beta ^2+100 r_{0}^4 \beta ^4+72 r_{0}^2 \beta ^6+16 \beta ^8}}{4 r_{0}^2 \beta ^{16}+4
\beta ^{18}}\bigg)^{1/2}
    \end{split}
\end{equation}
Interestingly, in bounding the parameter $A$ from expressions (\ref{eq471})--(\ref{eq48}), also are obtained some restrictions for the size of the wormhole throat $r_{0}$, namely: $\rho\geq 0 \Rightarrow r_{0}>0$, $\rho+p_{r}\geq 0\Rightarrow r_{0}>\sqrt{2\beta^{2}/3}$ and $\rho+p_{\perp}\geq 0\Rightarrow r_{0}>\sqrt{\beta^{2}/2}$. Recalling that all these bounds have been obtained by fixing $\alpha=2.0$. Furthermore, the conditions (\ref{Fcondition}) and (\ref{dencondition})--(\ref{wectcondition}) are the corresponding ones to simultaneously satisfy the expressions (\ref{eq471})--(\ref{eq48}) and violate the non--existence theorem (\ref{eq17}). The Figure \ref{fig7} illustrates the trend of the NEC and WEC at the wormhole throat and its neighborhood. As can be seen, in all cases these energy conditions are satisfied and, as the parameter $A$ decreases in magnitude (from left to right) the violation of these conditions is happening close to the throat. Then increasing $A$ in magnitude, of course respecting the previous discussion, the situation is more favorable. It should be noted that the condition $\rho+p_{\perp}$ is violated beyond the throat as it usually happens. Notwithstanding, as said above we are focused just at region compromising the throat $r_{0}$ and its vicinity $r_{0}+\epsilon$, being $\epsilon$ a positive small quantity. To reinforce the former arguments, we have plotted the space parameter validation of the energy conditions (NEC and WEC) and the violation of the non--existence theorem (see Figure \ref{fig8} and left panel of Figure \ref{fig9}). Moreover, the right panel of Figure \ref{fig9} is depicting the common region validation of all these conditions simultaneously.

So far, we have analyzed the possibility of having wormhole structures satisfying both, the NEC and WEC at the throat and its vicinity in the context of $f(R)$ gravity theory. The construction of this space--time is taking into account some engineering conditions to be a traversable tunnel connecting in this case two asymptotically flat regions. As a matter of fact only the flare--out condition have be taken into account. Of course, this is a necessary condition for a traversable wormhole but not enough. In the simplest case when $\Phi(r)=0$, as was done in \cite{Morris88}, tidal forces are not determinant concerning the traversability of the wormhole, at most one can determine or bound the travel velocity using the transverse tidal forces, since in the radial direction under the condition $\Phi(r)=0$ (or constant), there are not associated tidal forces. Nevertheless, when $\Phi(r)$ is incorporated, this quantity plays a major role in determining the traversability of the wormhole configuration. So, talking about traversable wormholes is non--trivial, since one must wonder if such a structure will be used for human travel or only to send some information through a beam of particles or light, for example. The next section is devoted to discuss this issue for the present model.   

\section{Humanely Traversable Wormholes?}\label{sec5.1}

In this section we shall discuss the main aspects and conditions for traversable wormholes with a non--vanishing red--shift. In general, traversability requires the fulfillment of some constraints on the red--shift function and its derivative, at the spatial stations location and the throat at the same time. These restrictions are \cite{Morris88,visser}\\

\textbf{Spatial station location}\\

At this point i.e., $l=l_{-}$ and $l=l_{+}$ where $l$ is the proper radial distance and $\pm$ stands for the upper and lower spatial station, respectively. The wormhole effects should be negligible, this is so because the spatial stations are far enough from the wormhole throat. Mathematically such conditions can be expressed as 
\begin{itemize}
    \item The space--time geometry must be nearly flat, then $b(r)/r<<1$.
    \item The gravitational red--shift of signals far away the station must be small i.e., $|\Phi(r)|<<1$.
    \item The measured gravitational acceleration at the stations, must be less than or of order 1 Earth gravity, $g_{\oplus}=9.81\, [\text{m}/\text{s}^{2}]$, thus $|\Phi'(r)|\lesssim g_{\oplus}$. However, it should be taken into account that in geometrized units ($c=G=1$) the numerical values of the Earth gravity acceleration is $g_{\oplus}=1.09\times 10^{-13}[\text{km}^{-1}]$.   \\
\end{itemize}

\textbf{Wormhole throat}\\

\begin{figure}[H]
\centering
\includegraphics[width=0.35\textwidth]{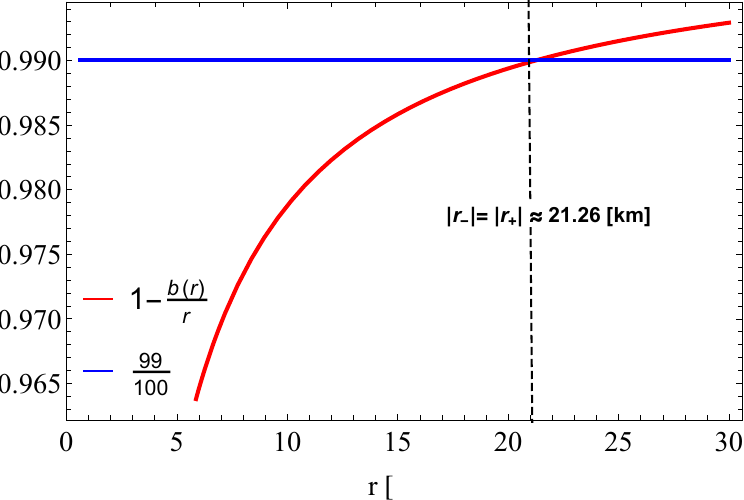} 
\caption{The radial location $\hat{r}=|r_{\pm}|$ of the spatial stations for values $A=3.0\,[\text{km}^{2}]$ and $\{r_{0};\beta\}=\{0.6;0.5\}\,[\text{km}]$. }
\label{fig10}
\end{figure}

\begin{itemize}
\item The full trip time should be less or of the order of 1 year approximately, as measured by the traveler and people living in the stations.
    \item The acceleration
felt by the radially moving traveler given by
\begin{equation}
    a(r)=\pm \left(1-\frac{b(r)}{r}\right)^{1/2}e^{-\Phi(r)}\left(\gamma e^{\Phi(r)}\right)^{\prime},
\end{equation}
where $\gamma=1/\sqrt{1-v^{2}(r)}$ and $v(r)$ being the radial velocity of the traveller. The acceleration $a(r)$ must not exceed by much 1 Earth gravity $g_{\oplus}$. That is
\begin{equation}
\bigg|\left(1-\frac{b(r)}{r}\right)^{1/2}e^{-\Phi(r)}\left(\gamma e^{\Phi(r)}\right)^{\prime}\bigg|\lesssim g_{\oplus}.
\end{equation}
\item The tidal accelerations exerted on the traveller must  
not exceed 1 Earth gravity $g_{\oplus}$. The tidal acceleration in the radial and lateral direction are
\begin{equation}\label{radial}
\left|\left(1-\frac{b(r)}{r}\right)\left(-\Phi^{\prime \prime}(r)-\Phi^{\prime 2}(r)+\frac{rb^{\prime}(r)-b}{2 r(r-b(r))} \Phi^{\prime}(r)\right) \right||\xi| \lesssim g_{\oplus},
\end{equation}
\begin{equation}\label{lateral}
\left|\frac{\gamma^{2} }{2 r^{2}}\left[\left(v(r)\right)^{2}\left(b^{\prime}(r)-\frac{b(r)}{r}\right)+2(r-b(r)) \Phi^{\prime}(r)\right]\right||\xi| \lesssim g_{\oplus},
\end{equation}
where $\xi$ corresponds to the traveller's size body.
As was pointed out in \cite{Morris88}, expression (\ref{radial}) can be seen as a constraint on the red--shift function, whilst (\ref{lateral}) can be regarded as constraining the speed $v(r)$ with which the traveler
crosses the wormhole.
\end{itemize}
Let us start studying the three first conditions on the spatial stations. To do so, first of all one needs to know the radial distance location of the stations. To obtain this information one can apply the same procedure as was done in \cite{Morris88}, that is, the factor $1-b(r)/r$ differs from unity at most by $1\%$. Applying this criterion and taking into account the numerical values $\{r_{0};\beta\}=\{0.6;0.5\}\, [\text{km}]$, one gets $\hat{r}=|r_{-}|=|r_{+}|\approx 21.26\,[\text{km}]$ (see Figure \ref{fig10}). Now, the radial proper distance $l$ can easily computed as follows
\begin{equation}
    l(r)=\pm \int^{\hat{r}}_{r_{0}}\frac{dr}{\sqrt{1-\frac{b(r)}{r}}}\approx \pm 21.47\,[\text{km}].
\end{equation}

\begin{figure}[H]
\centering
\includegraphics[width=0.34\textwidth]{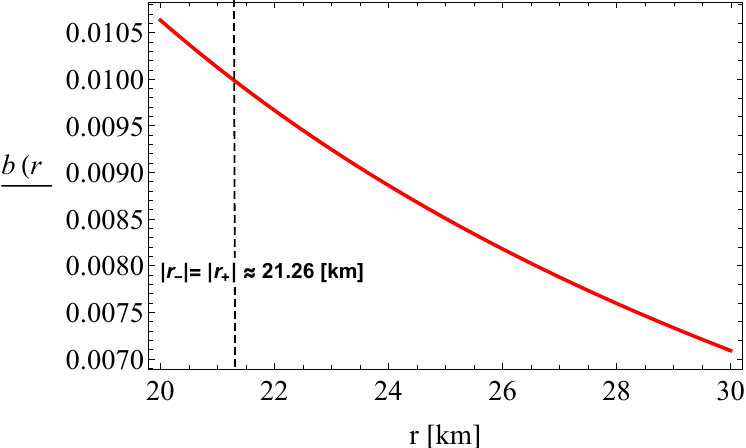} \ 
\includegraphics[width=0.31\textwidth]{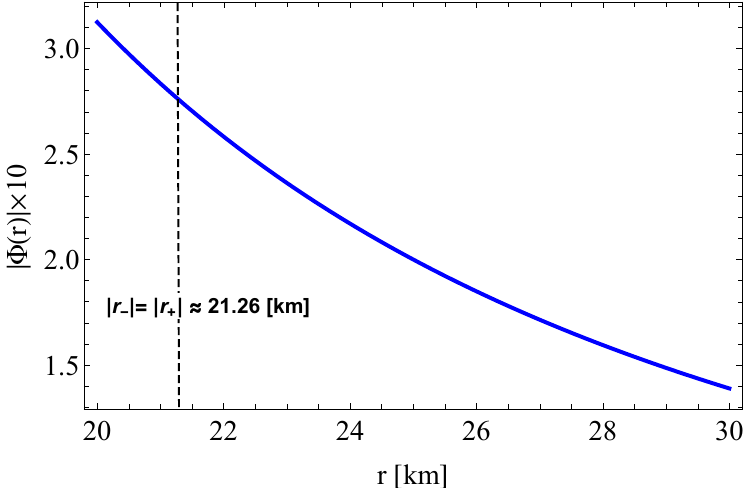}\
\includegraphics[width=0.31\textwidth]{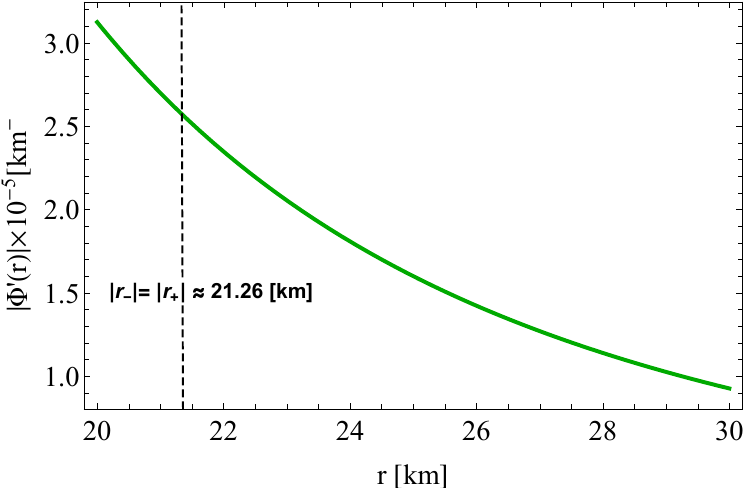} 
\caption{\textbf{Left panel}: The asymptotically flat condition at the spatial stations. \textbf{Middle panel}: The trend of the red--shift function at the spatial station versus the radial coordinate $r$. \textbf{Right panel}: The behavior of the derivative of the red--shift function at the stations against the radial coordinate $r$. To build these plots the following numerical values have been used $A=3.0\,[\text{km}^{2}]$ and $\{r_{0};\beta\}=\{0.6;0.5\}\,[\text{km}]$.     }
\label{fig11}
\end{figure}

It is clear that points $b(\hat{r})/\hat{r}<<1$ and $|\Phi(\hat{r})|<<1$ are satisfied at the spatial station location, however, $|\Phi^{\prime}(\hat{r})|\lesssim 1.09\times 10^{-3}\, [\text{km}^{-1}]$ is not. These conditions are displayed in Figure \ref{fig11}. Next, we can evaluate Equation (\ref{radial}) at the wormhole throat to get some information about the magnitude of the constant parameter $A$. Considering the human's body size $\xi=2\,[\text{m}]$, at the wormhole throat condition from (\ref{radial}) one obtains 
\begin{equation}\label{eq66}
    |\Phi^{\prime}(r)|\lesssim \frac{r}{|b^{\prime}(r)-1|}\times 10^{-10}[\text{km}^{-1}],
\end{equation}
providing for $\{r_{0};\beta\}=\{0.6;0.5\}\,[\text{km}]$ the following bound on the constant $A$
\begin{equation}\label{eq67}
    A\gtrsim 10^{10}.
\end{equation}

\begin{figure}[H]
\centering
\includegraphics[width=0.35\textwidth]{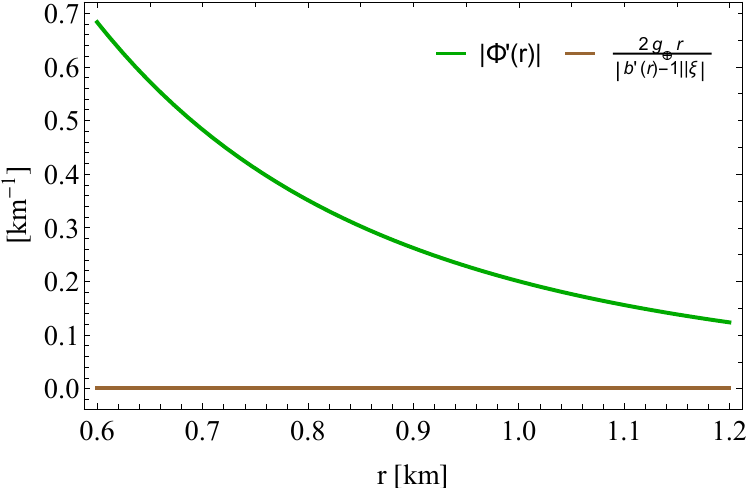} 
\caption{The trend of the gradient of the red--shift function (green line) and the right member of condition (\ref{eq66}) (brown line), versus the radial coordinate $r$. To build this plot the following numerical values have been used $A=3.0\,[\text{km}^{2}]$ and $\{r_{0};\beta\}=\{0.6;0.5\}\,[\text{km}]$.     }
\label{fig12}
\end{figure}

The above constraint is extremely far from the values
obtained trough expressions (\ref{flarecons}), (\ref{Fcondition}) and (\ref{dencondition})--(\ref{wectcondition}) (see below the summary in table \ref{table2}). Therefore, there is not possibility of having a human traversable wormhole space--time satisfying NEC and WEC. The trend of the gradient of the red--shift function and the right member of expression (\ref{eq66}) are exhibited in Figure \ref{fig12}. As can be seen, the gradient of the red--shift function is super--passing at the wormhole throat, the condition imposed by the right member of expression (\ref{eq66}).

\section{Conclusion}\label{sec6}
In this work, analytical wormhole solutions in the framework of $f(R)$ gravity theory have been explored. As building block to obtain the wormhole space--time geometry, the so--called class I condition (\ref{eq34}) has been employed. After a suitable choice of the red--shift function $\Phi(r)$ (\ref{eq36}) respecting all the requirements listed in Section \ref{sec2}, the form or shape function $b(r)$ was obtained by plugging (\ref{eq36}) into the Equation (\ref{eq34}), leading to (\ref{eq371}), after curing the wormhole throat condition $b(r_{0})=r_{0}$, and by adding a constant parameter $\delta$. This results in an asymptotically flat wormhole space--time, given by (\ref{eq381}). To corroborate this feature, the left panel of Figure \ref{fig1} illustrates the shape function trend and Figure \ref{fig3} displays the 2D and 3D embedding diagrams, exhibiting the usual asymptotically flat behavior. Furthermore, in left panel of Figure \ref{fig1} the flare--out condition for all $r$ is satisfied. 

Once the space--time geometry is specified, one needs to determine the energy--momentum tensor in order to analyze the energy conditions. Nevertheless, in the present situation, also it is necessary to fix the $f(R)$ gravity model. In this concern, we have chosen the model given by (\ref{eq49}). This model has good and viable cosmological properties. This $f(R)$ functional is depending on two parameters $\alpha$ and $\chi$. These parameters play an important role on the satisfaction of NEC and WEC. As the field Equations (\ref{eq9}) constitute an intricate set of equations, here we have evaluated these expressions at the wormhole throat to get some information about NEC and WEC at this point and its vicinity. This analysis leads to expressions (\ref{eq471})--(\ref{eq48}), where in order to meet these energy conditions, the $f(R)$ gravity model, its derivatives and the shape function must satisfy certain requirements. Besides, as it is well--known the violation of the non--existence theorem is a necessary condition in getting traversable wormholes satisfying at least NEC in the $f(R)$ gravity scenario. The Equations (\ref{eq9}) impose the constraints (\ref{conditionwecr}), (\ref{groned}) and (\ref{cond1}). Interestingly, the constraint (\ref{groned}) is bounding the behavior of the shape function $b(r)$. So, for the red--shift function (\ref{eq36}) the corresponding solution to the Gronwall--Bellman inequality, is obtained in the critical case. With this information at hand, through (\ref{groned}) the resulting critical shape function (\ref{shapegron}) should be bounded from below by (\ref{eq371}). This information can be corroborated in the right panel of Figure \ref{fig1}. On the other, for the chosen model (\ref{eq49}), Figures \ref{fig5} and \ref{fig6} are confirming the violation of the non--existence theorem (right panel) and conditions (\ref{conditionwecr}) (left panel) and (\ref{cond1}) (right panel), respectively. This analysis allowed to restrict the constant parameter $A$, obtaining the bounds (\ref{flarecons}), (\ref{Fcondition}) and (\ref{dencondition})--(\ref{wectcondition}). In table \ref{table2} are summarizing the numerical values for $A$ taking into account some numerical values for the parameters $\{r_{0};\beta;\chi;\alpha\}$. In Figure \ref{fig7} it is evident that NEC and WEC are satisfied at the wormhole throat and its neighborhood. Additionally. Figures \ref{fig8} and \ref{fig9} (left panel) are exhibiting the validation regions for energy conditions and violation of non--existence theorem for the proposed toy model, what is more the right panel of Figure \ref{fig9} is illustrating the common region (purple region) where all these conditions are simultaneously satisfied.  

\begin{table}[H]
\begin{center}
\resizebox{17cm}{!} {\begin{tabular}{|c | c |c|c| }
\hline
\multicolumn{4}{ |c| }{Bounds on parameter $A$ at the wormhole throat} \\ \hline
$\rho+p_{r}\geq 0$, \ $\rho+p_{\perp}\geq 0$ & $\rho\geq 0$, \   $\rho+p_{r}\geq 0$, \ $\rho+p_{\perp}\geq 0$ & $F<0$ & $b^{\prime}$  \\ \hline
 $A\geq 0.718$, \ $0.254\leq A\leq 3.151$ & $A\geq 1.021, A\geq 0.718$, \ $0.254\leq A\leq 3.151$ &$0.054<A<5.181$ & $A>-0.285$  
\\ \hline
\end{tabular}}
\caption{The numerical constraints imposed on $A$ by the energy conditions, violation of the non--existence theorem and flare--out condition, for $\chi=2.5\,[\text{km}^{-2}]$, $\{r_{0};\beta\}=\{0.6;0.5\}\,[\text{km}]$ and $\alpha=2.0$. }
\label{table2}
\end{center}
\end{table}

Finally, we have explored the possibility of having human traversable wormhole solutions, satisfying NEC and WEC in the $f(R)$ gravity background. In this concern, the model is representing a traversable wormhole space--time, but not for human travel. This is so because, the radial tidal acceleration at the wormhole throat, introduced by the red--shift function, exceeded 1 Earth gravity acceleration $g_{\oplus}$. In fact, from (\ref{radial}) evaluated at the throat, one obtains a constraint (\ref{eq66}) for the gradient of the red--shift. This condition leads to a bound on the constant $A$ (see Equation (\ref{eq67})), which does not match with those values shown in table \ref{table2}, which simultaneously satisfy the NEC, WEC, flare--out condition and violation of the non--existence theorem. Moreover, as in shown Figure \ref{fig11} traversability conditions are not satisfied at the spatial station position too, since the gradient of the red--shift function at this point overcomes 1 Earth gravity acceleration. The same situation occurs at the wormhole throat as illustrated in Figure \ref{fig12}, where the condition (\ref{eq66}) is not fulfilled. Notwithstanding, these restrictions do not preclude the possibility of having wormhole solutions available for human travel while satisfying the NEC and WEC at the throat and its neighborhood in the $f(R)$ gravity scenario. Of course, this could be merely a drawback of either the model geometry or the $f(R)$ function, or perhaps an inappropriate combination of both ingredients. Therefore, it would be ideal to investigate in a completely general way, under what conditions wormholes available for a hypothetical interstellar journey manned by humans is possible within the framework of $f(R)$ gravity theory. As final comment, it is clear that there are still several aspects that could be investigated in the future (not only the aforementioned). For instance, one may study the stability of the solution. This issue shall be addressed elsewhere. 

\appendix

\section{Deriving the class I condition}\label{A}

In this appendix, necessary and sufficient conditions that any spherically symmetric space--time (static and non--static) must satisfy to be a class I space--time, are derived in details. The starting point is to consider the Riemannian curvature in terms of the extrinsic curvature or second fundamental form (Gauss's equation) in conjunction with the Codazzi's equation. So, these conditions read 
\begin{itemize}
    \item A system of symmetric quantities $b_{\mu\,\nu}$ must be established, such that
    \begin{equation}\label{eq19}
    R_{\mu\,\nu\,\alpha\,\beta}=\epsilon\,\left(b_{\mu\,\alpha}\,b_{\nu\,\beta}-b_{\mu\,\beta}\,b_{\nu\,\alpha}\right)\quad \left(\text{Gauss's equation}\right),
    \end{equation}
    where $\epsilon=\pm1$ whenever the normal to the manifold is space--like (+1)
or time--like (-1).
\item The system $b_{\mu\nu}$ must satisfy the differential equations
\begin{equation}\label{eq20}
\nabla_{\alpha}b_{\mu\,\nu}-\nabla_{\nu}\,b_{\mu\,\alpha}=0 \quad \left(\text{Codazzi's equation}\right).
\end{equation}
\end{itemize}
From 
\begin{equation}\label{eq188}
ds^{2}=-e^{\eta(r)}dt^{2}+e^{\lambda(r)}dr^{2}+r^2(d\theta^2+sin^2\theta d\phi^2)
\end{equation}
the non vanishing elements of the Riemann tensor are
 \begin{eqnarray}\nonumber
 R_{trtr}&=&-\frac{e^{\eta}}{4}\left[2\,\eta^{\prime\prime}+\eta^{\prime \ 2}-\lambda^{\prime}\,\eta^{\prime}\right], \\ \nonumber
 R_{\theta\,\phi\,\theta\,\phi}&=& -e^{-\lambda}\,r^{2}\,sin^{2}\theta\left(e^{-\lambda}-1\right), \\ \label{eq21}
 R_{r\,\phi\, r\,\phi}&=&sin^{2}\theta\, R_{r\,\theta \,r\,\theta}=\frac{r}{2}\,\lambda^{\prime}, \\ \nonumber
 R_{t\,\theta\, t\,\theta}&=&sin^{2}\theta \,R_{t\,\phi\, t\,\phi}=-\frac{r}{2}\,\eta^{\prime}\,e^{\eta-\lambda}.
 \end{eqnarray}
So, by using the set of Equations (\ref{eq21}) into Equation (\ref{eq19}) one gets
\begin{eqnarray}&&\label{eq22}
b_{tr}\,b_{\phi\,\phi}=R_{r\phi t\phi}=0;~~
b_{tr}\,b_{\theta\,\theta}=R_{r\theta t\theta}=0;\\ \nonumber&&
b_{tt}\,b_{\phi\,\phi}=R_{t\phi \,t\,\phi};~~
b_{tt}\,b_{\theta\,\theta}=R_{t\theta t\theta}; ~~
b_{rr}\,b_{\phi\,\phi}=R_{r\phi r\phi}; \\ \nonumber&&
b_{\theta\,\theta}\,b_{\phi\,\phi}=R_{\theta\,\phi\,\theta\,\phi}; ~~
b_{rr}\,b_{\theta\,\theta}=R_{r\theta\, r\theta}; ~~
b_{tt}\,b_{rr}=R_{trtr}.
\end{eqnarray} 

The above relations leads to:
\begin{equation}\label{eq23}
\begin{split}
\left(b_{tt}\right)^{2}=\frac{\left(R_{t\,\theta\, t\,\theta}\right)^{2}}{R_{\theta\,\phi\,\theta\,\phi}}\,sin^{2}\theta, \quad \left(b_{rr}\right)^{2}=\frac{\left(R_{r\,\theta \,r\,\theta}\right)^{2}}{R_{\theta\,\phi\,\theta\,\phi}}\,sin^{2}\theta, \quad \left(b_{\theta\,\theta}\right)^{2}=\frac{R_{\theta\,\phi\,\theta\,\phi}}{sin^{2}\theta}, \quad \left(b_{\phi\,\phi}\right)^{2}=sin^{2}\theta \,R_{\theta\,\phi\,\theta\,\phi}. \end{split}
\end{equation}
Upon replacing (\ref{eq23}) into  expression (\ref{eq22}) one gets
\begin{equation}\label{eq24}
R_{t\,\theta\, t\,\theta}\,R_{r\,\phi\, r\,\phi}=R_{trtr}\,R_{\theta\,\phi\,\theta\,\phi},    
\end{equation}
subject to $R_{\theta\,\phi\,\theta\,\phi}\neq\,0$ \cite{sharma}. It should be noted that Equation (\ref{eq22}) satisfies Codazzi's Equation (\ref{eq20}). On the other hand, in the case of a general static spherically symmetric space--time, the second and last equality in (\ref{eq23}) becomes
\begin{equation}\label{eq25}
b_{tr}\,b_{\theta\,\theta}=R_{r\,\theta \,t\,\theta} \quad \mbox{and}\quad  b_{tt}\,b_{rr}-\left(b_{tr}\right)^{2}=R_{trtr},  
\end{equation}
where $\left(b_{tr}\right)^{2}=sin^{2}\theta \left(R_{r\theta t \theta}\right)^{2}/R_{\theta\,\phi\,\theta\,\phi}$. So, the class I condition becomes \cite{eisland,karmarkar}
\begin{equation}\label{eq26}
R_{t\,\theta\, t\,\theta}\,R_{r\,\phi\, r\,\phi}=R_{trtr}\,R_{\theta\,\phi\,\theta\,\phi}+R_{r\theta t \theta}\,R_{r\,\phi\, t\,\phi}.    
\end{equation}
In this particular case, where the space--time is given by Equation (\ref{eq188}) the condition (\ref{eq23}) (or equivalently (\ref{eq25})) leads to
\begin{equation}\label{eq27}
2\frac{\eta^{\prime\prime}}{\eta^{\prime}}+\eta^{\prime}=\frac{\lambda^{\prime}\,e^{\lambda}}{e^{\lambda}-1},   
\end{equation}
with $e^{\lambda}\neq 1$. This equation can be solved to express $\eta(r)=\eta\left(\lambda(r)\right)$ or $\lambda(r)=\lambda\left(\eta(r)\right)$. The results are:
\begin{equation}\label{eq28}
e^{\lambda}=1+A\,\eta^{\prime \, 2}\,e^{\eta},    
\end{equation}
or
\begin{equation}\label{eq29}
e^{\eta}=\left[B+C\int\,\sqrt{\left(e^{\lambda}-1\right)}\,dr\right]^{2},    
\end{equation}
being $\{A,B,C\}$ integration constants.\\

\section*{Acknowledgement} 
A. S. A. acknowledges the financial support provided by University Grants Commission (UGC) through Senior Research Fellowship (File No. 16-9 (June 2017)/2018 (NET/CSIR)),  to carry out the research work. F. Tello-Ortiz thanks the financial support by projects ANT--1956 and SEM 18--02 at the Universidad de Antofagasta, Chile. F. Tello-Ortiz acknowledges
the PhD program Doctorado en Física mención en Física Matemática de la Universidad de Antofagasta for continuous support and encouragement. B. M. acknowledges IUCAA, Pune (India) for hospitality and support during an academic visit where a part of this work is accomplished.

\end{document}